\documentclass[twocolumn]{aastex701}

\usepackage{amsmath}
\usepackage[version=4]{mhchem}
\usepackage{hyperref}

\begin{document}

\title{JWST Transmission Spectroscopy of TOI-3235 b: Challenges in Constraining Giant Planet Atmospheres Around M Dwarfs Amid Stellar Contamination}

\author[0000-0002-4962-2543]{Zoe Ko}
\affil{William H. Miller III Department of Physics and Astronomy, Johns Hopkins University, Baltimore, MD 21218, USA}
\email{zko2@jh.edu}

\author[0000-0001-9513-1449]{N\'estor Espinoza}
\affil{Space Telescope Science Institute, 3700 San Martin Drive, Baltimore, MD 21218, USA}
\affil{William H. Miller III Department of Physics and Astronomy, Johns Hopkins University, Baltimore, MD 21218, USA}
\email{nespinoza@stsci.edu}

\author[0000-0002-5389-3944]{Andrés Jordán}
\affil{Facultad de Ingenier\'ia y Ciencias, Universidad Adolfo Ib\'{a}\~{n}ez, Av. Diagonal las Torres 2640, Pe\~{n}alol\'{e}n, Santiago, Chile}
\affil{Departamento de Astronom\'ia, Universidad de Chile, Casilla 36-D, Santiago, Chile}
\email{ajordan@astrofisica.cl}

\author[0000-0001-6023-1335]{Daniel Bayliss}
\affil{Department of Physics, University of Warwick, Gibbet Hill Road, Coventry CV4 7AL, UK}
\affil{Centre for Exoplanets and Habitability, University of Warwick, Gibbet Hill Road, Coventry CV4 7AL, UK}
\email{d.bayliss@warwick.ac.uk}

\author[0000-0001-7904-4441]{Edward M. Bryant}
\affil{Department of Physics, University of Warwick, Gibbet Hill Road, Coventry CV4 7AL, UK}
\affil{Centre for Exoplanets and Habitability, University of Warwick, Gibbet Hill Road, Coventry CV4 7AL, UK}
\email{edward.m.bryant@warwick.ac.uk}

\author[0000-0001-8355-2107]{Martin Schlecker}
\affil{European Southern Observatory, Karl-Schwarzschild-Straße 2, 85748 Garching bei M\"unchen, Germany}
\email{martin.schlecker@eso.org}

\author[0000-0002-5945-7975]{Melissa J. Hobson}
\affil{Observatoire de Genève, Département d'Astronomie, Université de Genève, Chemin Pegasi 51b, 1290 Versoix, Switzerland}
\email{melissa.hobson@unige.ch}

\author[0000-0002-9158-7315]{Rafael Brahm}
\affil{Facultad de Ingenier\'ia y Ciencias, Universidad Adolfo Ib\'{a}\~{n}ez, Av. Diagonal las Torres 2640, Pe\~{n}alol\'{e}n, Santiago, Chile}
\affil{Millennium Institute for Astrophysics, Nuncio Monse\~{n}or Sotero Sanz 100, Of. 104, Providencia, Santiago, Chile}
\email{rafael.brahm@uai.cl}

\author[0000-0002-9020-7309]{Remo Burn}
\affil{Laboratoire Lagrange, UMR7293, Universit\'e C\^ote d'Azur, CNRS, Observatoire de la C\^ote d'Azur, Boulevard de l’Observatoire, 06304, Nice Cedex 4, France}
\affil{Max Planck Institute for Astronomy, K\"onigstuhl 17, 69117 Heidelberg, Germany}
\email{remo.burn@oca.eu}

\author[0000-0002-1493-300X]{Thomas Henning}
\affil{Max Planck Institute for Astronomy, K\"onigstuhl 17, 69117 Heidelberg, Germany}
\email{henning@mpia.de}

\begin{abstract}
Giant planets orbiting low-mass M dwarfs challenge current planet formation theories, which predict that such planets are unlikely to form. Characterizing their atmospheres can provide key insight into their origins, but analysis is complicated by stellar contamination in transmission spectra. We present a JWST/NIRSpec PRISM transmission spectrum of TOI-3235~b, a 604~K, $0.665~M_{J}$ giant planet orbiting a $0.39\,M_\odot$ M dwarf. We apply a hierarchical retrieval framework incorporating an atmospheric model, a parametric stellar contamination model, and a Gaussian process (GP) to capture residual structure not explained by these deterministic components. We find that the inferred atmospheric properties are model-dependent: retrievals that treat stellar contamination deterministically yield a constrained CH$_4$ abundance corresponding to a sub-solar metallicity that would challenge standard expectations from both core accretion and gravitational instability, and potentially elevated CO and CO$_2$ abundances that would require chemical disequilibrium. However, residual wavelength-correlated structure suggests the deterministic model is incomplete. Including a GP to marginalize over this structure broadens the atmospheric posterior distributions, preventing robust constraints on the atmospheric composition and highlighting the challenge of characterizing giant planet atmospheres around M dwarfs when physical models are incomplete. We demonstrate that an eclipse observation could clarify these model-dependent ambiguities: detected emission features would constrain the planet's metallicity and test formation scenarios, while a featureless spectrum would confirm stellar contamination dominates the transmission spectrum, providing an empirical M dwarf contamination spectrum.
\end{abstract}


\section{Introduction} \label{sec:intro}

The prevailing theory of giant planet formation points to core accretion \citep{pollack1996, alibert2005, liu2020}, which predicts that giant planets are unlikely to form around stars with masses below $M_* \lesssim 0.5~M_\odot$. Disk mass is expected to scale with host star mass \citep{andrews2013, pascucci2016, tychoniec2020, mulders2021, manara2022}, causing low-mass M dwarfs to have less material to form a planetary core massive enough (\(\sim 10\,M_\oplus\)) to initiate runaway gas accretion \citep{stevenson1982, pollack1996}. Additionally, low masses lead to longer Keplerian orbital periods, which further prevent sufficient material from building up before the disk dissipates \citep{laughlin2004}. These theories are further supported by modern state-of-the-art formation models; for example, \citealt{burn2021} find that no planets with masses $M_p > 100\,M_\oplus$ form around stars with masses $M_* < 0.5\,M_\odot$ in their standard models. Similar results are found in the context of pebble accretion models by, e.g., \citealt{liu2019} and \citealt{chachan2023}. 

Despite these theoretical predictions, empirical observations have proven otherwise, challenging existing formation models. Radial velocity observations have demonstrated that Jupiter-mass planets with short orbital periods (1--10 days) are rare but do occur around M dwarfs \citep{bonfils2013, pinamonti2022, ribas2023, pass2023, mignon2025}, and TESS has characterized over 30 giant planets on short orbits around M dwarfs, with the ones having the lowest mass host stars ($M_* \lesssim 0.4 M_\odot$) being: TOI-6894b ($M_* = 0.207 \pm 0.011~M_\odot$) \citep{bryant2025}, TOI-4860b ($M_* = 0.3357 \pm 0.0081~M_\odot$) \citep{almenara2024}, TOI-3235~b ($M_* = 0.3939 \pm 0.0200~M_\odot$) \citep{hobson2023}, TOI-5205b ($M_* = 0.392 \pm 0.015~M_\odot$) \citep{kanodia2023}, and TOI-3884b ($M_* = 0.2813 \pm 0.0067~M_\odot$) \citep{almenara2022}. These discoveries have spurred a new set of hypotheses that modify traditional core accretion theories, such as incorporating disk metallicity effects. An increased disk metallicity can promote core growth by providing a larger reservoir of solids \citep{laughlin2004, burn2021}, and models show that increasing the disk metallicity by a factor of ten can lead to an increase of more than three orders of magnitude in planet mass \citep{liu2019}. This mass-metallicity trend is well established for FGK-type stars, making it plausible that the same relationship could extend to M dwarfs, with high metallicities enabling giant planet formation. Structures in the protoplanetary disk can also act as migration traps \citep{masset2006}, stalling Type~I disk migration and enabling growing planetary cores to accrete more material \citep{burn2021, schlecker2022}. Formation may also occur in disks that are unusually massive; population synthesis models suggest that disk masses at the upper end of the observed distribution are necessary to reproduce the population of giant planets around low-mass stars \citep{burn2021}. Early formation during the disk's lifetime is another possibility, as planet formation that begins within the first Myr would have a longer timescale before significant disk dissipation occurs \citep{delamer2024}. Finally, a different alternative to the above-mentioned models would be gravitational instability, which could provide pathways for giant planet formation around M dwarfs \citep[see, e.g.,][]{boss2023}. However, this requires large disk-to-star mass ratios, which are relatively rare for low-mass stars \citep{rafikov2005can, andrews2007high, mercer2020}, and give rise to more massive and distant planets than those discovered so far \citep{kratter2010, schib2025}.

Detailed atmospheric characterization offers a powerful approach to probe initial formation channels, as atmospheric metrics such as metallicity and elemental ratios (e.g. C/O ratio) provide empirical diagnostics of planetary formation history and conditions, especially for gas giant exoplanets \citep{oberg2011, mordasini:2016, espinoza:2017,madhusudhan2019, molliere2022}. The \textit{James Webb Space Telescope} (\textit{JWST}) has become the premier facility for constraining these key parameters from exoplanet atmospheric observations, as its wide wavelength coverage allows it to constrain the abundances of key carbon and oxygen-bearing molecules \citep[see, e.g.,][for a review]{EP:2025}. However, extracting reliable atmospheric properties from transmission spectra of planets orbiting M dwarfs poses significant observational challenges, with the principal limitation being stellar contamination. 

By measuring wavelength-dependent changes as starlight filters through a planet's limb during transit \citep{seager2000theoretical}, transmission spectroscopy is fundamentally limited by the nature of the host star. Active regions such as star spots and faculae can introduce features into the transmission spectrum, imprinting signatures that may mimic or obscure planetary atmospheric characteristics \citep{rackham2018}. M dwarf systems are particularly susceptible to the effects of stellar contamination, with higher levels of activity and spot-covering fractions compared to earlier spectral types \citep{rackham2018, somers2020spots}. Physically modeling stellar contamination is challenging; existing models are often overly simplified \citep{witzke2022can, garcia2022hst, norris2023spectral, lim2023atmospheric, rackham2024toward} or too computationally expensive to implement into current retrieval frameworks \citep{smitha2024first}. Beyond physical models, Gaussian processes (GPs) offer a promising statistical approach to model residual sources of wavelength-correlated structure in transmission spectra of exoplanets \citep[see, e.g.,][]{guilluy2024gaps, McCreery:2025, rotman2025enabling}. \citealt{espinoza2025dreams} showed that they can be used to mitigate the effects of stellar contamination from stars as active as TRAPPIST-1.

% Typically, stellar contamination models approximate spots and faculae with cooler and hotter 1D stellar spectra and combine these with the photospheric component, weighted by their respective surface covering fractions \citep{rackham2018}. However, these stellar spectra models \citep[e.g.][]{husser2013new, allard2013bt} do not capture the full complexity of stellar heterogeneities, excluding effects such as magnetic fields and limb-dependent contrasts, both of which influence the inferred spectral features \citep{witzke2022can, garcia2022hst, norris2023spectral, lim2023atmospheric, rackham2024toward}. Advances using 3D radiative magnetohydrodynamic simulations further reveal discrepancies between simplified 1D models and the actual impact of spots and faculae on the emergent spectrum \citep{smitha2024first}; however, the computational expense of detailed 3D simulations makes them impractical for implementation in current atmospheric retrieval frameworks.

Here, we present a detailed \textit{JWST} atmospheric characterization of one of these unlikely giant planets orbiting an M dwarf: TOI-3235~b, a 604~K $0.665~M_{J}$ gas giant orbiting a $0.39\,M_\odot$ M dwarf \citep{hobson2023}. This planet was observed as part of a Cycle~2 program targeting giant planets around low-mass stars, with the goal of characterizing their atmospheric properties to shed light on their formation mechanisms (GO 3731, PI: Jord\'an). As we detail below, the dataset exhibits strong signatures of stellar contamination, which we address with the GP-aided retrieval framework following \citealt{espinoza2025dreams}, enabling a robust investigation of the planet's atmospheric properties.

We present our observations and data reduction in Section~\ref{sec:data}, our atmospheric retrieval set-up in Section~\ref{sec:retrievals}, and our results in Section~\ref{sec:results}. We discuss implications of our results in Section~\ref{sec:discussion} and conclude in Section~\ref{sec:conclusion}.

\section{Observations and Data Reduction} \label{sec:data}

\subsection{Observing Program} \label{subsec:observation}
We observed one primary transit of TOI-3235~b on July 27, 2024 as part of GO 3731 (PI: Jord\'an) using NIRSpec/PRISM with the SUB512S (512 x 16 pixel) subarray, which covers wavelengths from $0.63~\mu\mathrm{m}$ to $5.59~\mu\mathrm{m}$. We obtained 25,441 integrations of 0.74 seconds each, with four groups per integration, resulting in a total exposure time of 5.2296 hours, which comfortably covers our $T_{14}$ = 1.4796 hour-transit event.

\subsection{Data Reduction} \label{subsec:reduction}
We use \texttt{transitspectroscopy} \citep{transitspectroscopy} to perform data reduction, which draws on the JWST Calibration Pipeline \citep{bushouse2023jwst} to reduce the uncalibrated FITS files into the rates per integration. We follow the \texttt{transitspectroscopy} reduction process as described in \citealt{espinoza2025dreams}, Appendix A.1, which produces a flux time series for every wavelength in the NIRSpec/PRISM range.

We identify three pixels in the 2-D spectrum with abnormally high values at pixel coordinates [6,125], [6,76], and [4,421], most likely due to cosmic ray hits that were missed by the calibration pipeline. We therefore exclude these three pixels from our analysis.

\begin{figure}
    \centering
    \includegraphics[width=\linewidth]{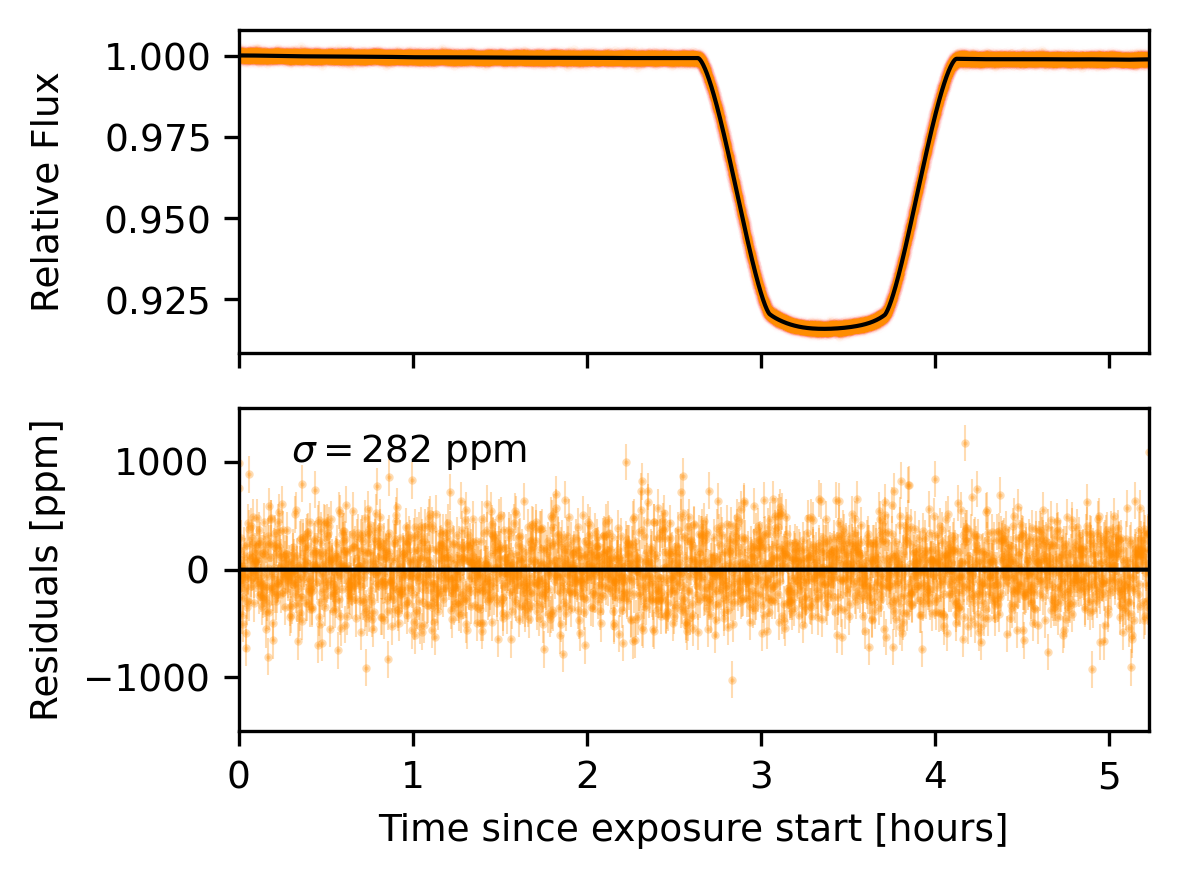}
    \caption{White light curve fit against data (top panel) and residuals (bottom panel).
}
    \label{fig:whitelight_fit}
\end{figure}

We fit transit light curves using \texttt{juliet} \citep{juliet}. While some studies report the shortcomings of a quadratic limb-darkening law and recommend using higher order prescriptions \citep[e.g.][]{mercier2025, coulombe2024biases}, we compare a three-parameter law with a quadratic law \citep{kipping2013efficient} and find that the two models have negligible differences in the produced transmission spectra. Therefore, we adopt the quadratic limb-darkening law to minimize model complexity. We first fit the band-integrated, white light curve to obtain updated transit parameters for our system by adding up the flux from all wavelengths. To model instrumental systematics, we fit a flux offset to account for light curve normalization, a jitter term to model possible underestimated error bars, and a linear term and GP term to model instrumental or astrophysical systematics. The priors and posteriors of this fit are presented in Table~\ref{tab:lightcurve}; the best-fit model is presented in Figure~\ref{fig:whitelight_fit}. No occulted spot-crossing events were observed in the light curves.

\begin{table*}[t!]
\centering
\caption{
Priors and posteriors for the white light curve fit for TOI-3235~b.
$N(\mu, \sigma^2)$ denotes a normal prior, $U(a, b)$ a uniform prior, log\,$U(a, b)$ a log-uniform prior, and ``fixed'' a fixed value.
}
\label{tab:lightcurve}
\begin{tabular}{llcc}
\hline
Parameter & Description & Prior & Posterior \\
\hline
\multicolumn{4}{c}{Physical \& orbital parameters} \\
\hline
$P$ [days]           & Orbital period          & $N(2.6,\,0.1^2)^{\mathrm{a}}$                       & $2.618^{+0.065}_{-0.066}$ \\
$t_0$ [BJD]          & Transit center time     & $U(2460519.498,\,2460519.716)^{\mathrm{b}}$          & $2\,460\,519.639327^{+0.000009}_{-0.000010}$ \\
$R_p/R_*$            & Planet-to-star radius ratio           & $U(0,\,1)$                                           & $0.2817^{+0.00033}_{-0.00035}$ \\
$b$                  & Impact parameter        & $N(0.51,\,0.01^2)^{\mathrm{a}}$                      & $0.486^{+0.0015}_{-0.0015}$ \\
$a/R_*$              & Scaled semi-major axis  & $N(15.75,\,0.73^2)^{\mathrm{a}}$                     & $15.53^{+0.39}_{-0.39}$ \\
$e$                  & Eccentricity           & fixed (0.03)$^{\mathrm{a}}$                          & fixed (0.03) \\
$\omega$ [deg]       & Argument of periastron  & fixed (90)$^{\mathrm{a}}$                             & fixed (90) \\
\hline
\multicolumn{4}{c}{Quadratic limb-darkening coefficients} \\
\hline
$u_1$                &                        & $U(-3,\,3)$                                          & $0.127^{+0.035}_{-0.034}$ \\
$u_2$                &                        & $U(-3,\,3)$                                          & $0.204^{+0.058}_{-0.056}$ \\
\hline
\multicolumn{4}{c}{Instrument systematics} \\
\hline
$M_\mathrm{dilution}^{\mathrm{c}}$ & Dilution factor             & fixed (1)                                    & fixed (1) \\
$M_\mathrm{flux}^{\mathrm{c}}$     & Relative flux offset        & $N(0,\,0.1^2)$                               & $0.00066^{+0.000032}_{-0.000037}$ \\
$\sigma_w$ [ppm]                   & White noise/jitter term     & log\,$U(10,\,1000)$                          & $227.9^{+4.8}_{-4.9}$ \\
$\theta_0^{\mathrm{d}}$            & Linear baseline slope       & $U(-10,\,10)$                                & $-0.00029^{+0.000028}_{-0.000030}$ \\
$\sigma_{\mathrm{GP}}^{\mathrm{e}}$& GP amplitude                & log\,$U(10^{-6},\,10^{-2})$                  & $0.000105^{+0.000033}_{-0.000021}$ \\
$\rho_{\mathrm{GP}}$ [days]$^{\mathrm{e}}$ & GP timescale       & log\,$U(10^{-5},\,10^{3})$                   & $0.0079^{+0.0037}_{-0.0023}$ \\
\hline
\end{tabular}
\vspace{0.5em}

{\footnotesize
\begin{tabular}{@{}l@{}}
\textbf{Notes.}\\
$^{\mathrm{a}}$ Prior values adopted from \citet{hobson2023}.\\
$^{\mathrm{b}}$ Prior range reflects the start and end of the observing window.\\
$^{\mathrm{c}}$ See Section 2.1 of \citet{juliet} for a description of parameters.\\
$^{\mathrm{d}}$ Linear regressor used to capture linear trends in the data.\\
$^{\mathrm{e}}$ Mat\'ern~3/2 kernel GP implemented with \texttt{celerite} \citep{celerite} and used to capture systematics.
\end{tabular}
}

\end{table*}

\begin{figure}
    \centering
    \includegraphics[width=\linewidth]{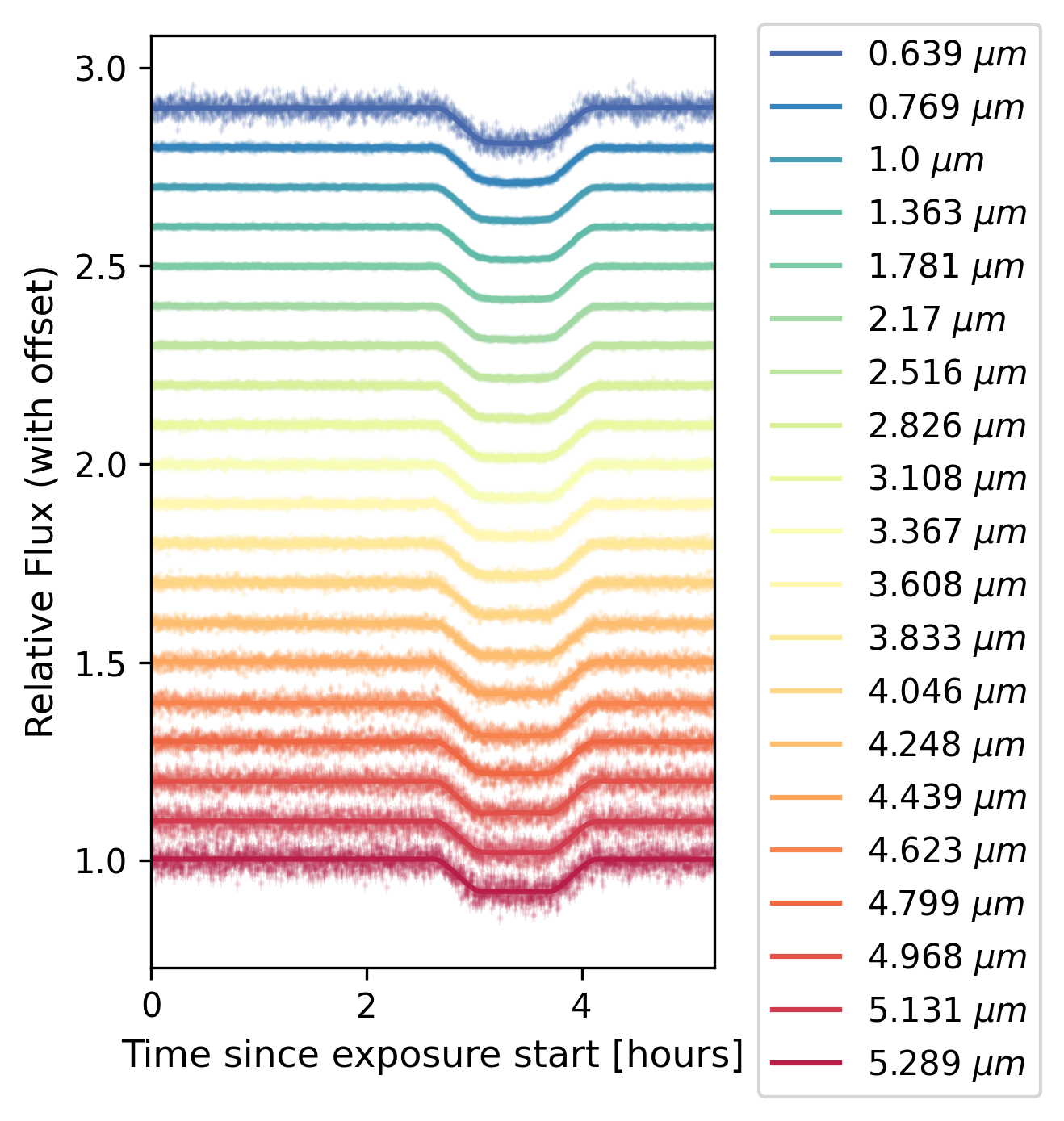}
    \caption{Sample wavelength-dependent light curve fits. There are a total of 441 curves spanning from $0.63~\mu\mathrm{m}$ to $5.59~\mu\mathrm{m}$.}
    \label{fig:wl_dep_fit}
\end{figure}

To fit the wavelength-dependent light curves, we fix all system parameters to those obtained from the white light curve analysis, allowing only the planet-to-star radius ratio, limb-darkening coefficients and instrumental systematics to vary with wavelength. Figure~\ref{fig:wl_dep_fit} shows examples of wavelength-dependent light curve fits across the spectral range.

We obtain the transmission spectrum through the wavelength-dependent light curve fits and bin the transmission spectrum to a resolution of $R \equiv \lambda / \Delta\lambda = 100$ to reduce computational demand for atmospheric retrievals, which we describe next.

\section{Atmospheric Retrievals} \label{sec:retrievals}
We follow the atmospheric retrieval framework as described in \citealt{espinoza2025dreams} Appendix C, using \texttt{POSEIDON} \citep{macdonald2017, macdonald2023poseidon} to conduct a Bayesian inference on the transmission spectrum of TOI-3235~b. We use \texttt{dynesty}'s Dynamic Nested Sampler \citep{speagle2020dynesty}, with the \texttt{rwalk} sampling algorithm and an initial number of $5{,}000$ live points to sample our prior space. We adopt broad, uninformative priors to allow full exploration of the parameter space and minimize potential biases. The priors used in our retrievals are summarized in Table~\ref{tab:priors}.

\begin{table*}[t!]
\centering
\caption{Retrieval parameters and prior bounds.}
\label{tab:priors}
\begin{tabular}{lccc}
\hline
Parameter & Lower Bound & Upper Bound & Prior Type \\
\hline
\multicolumn{4}{c}{Atmospheric Model Parameters} \\
\hline
Atmospheric Temperature $T$ [K] & $10^2$ & $10^3$ & Uniform \\

Log Volume Mixing Ratios of Trace Species $\log X$ & $-12$ & $-1$ & Uniform \\
\multicolumn{4}{l}{
\hspace{1em}
\ce{CO2}, \ce{CH4}, \ce{H2O}, \ce{NH3}, \ce{HCN}, \ce{CO},
\ce{SO2}, \ce{H2S}, \ce{C2H2}, \ce{C2H4}
} \\
Log Cloud Top Pressure $\log P_{\rm cloud}$ [bar] & $-6$ & $3$ & Uniform \\
Log Haze Amplitude $\log a$ & $-3$ & $8$ & Uniform \\
Haze Scattering Slope $\gamma$ & $-30$ & $30$ & Uniform \\
Log Reference Pressure $\log P_{\rm ref}$ [bar] & $-6$ & $3$ & Uniform \\
Log White Noise Amplitude $\log \sigma_{\rm w}$ & $-5$ & $1$ & Uniform \\
Flux Offset [ppm] & $-10^4$ & $10^4$ & Uniform \\
\hline
\multicolumn{4}{c}{Stellar Contamination Model Parameters} \\
\hline
Photosphere Temperature $T_{\rm phot}$ [K] & 3388 (Mean) &  68 (SD) & Normal \\
Stellar Heterogeneity Temperature(s) $T_{\rm het}$ [K] & 1200 & 7000 & Uniform \\
Log Stellar Heterogeneity Covering Fraction(s) $\log f_{\rm het}$ & $-3$ & $0$ & Uniform \\
\hline
\multicolumn{4}{c}{Gaussian Process Hyperparameters} \\
\hline
Log GP Length Scale $\log \rho$ [$\mu$m] & $-3$ & $2$ & Uniform \\
Log GP Amplitude $\log \sigma_{\rm GP}$ & $-5$ & $1$ & Uniform \\
\hline
\end{tabular}
\end{table*}

Each retrieval is a combination of one or more of the following three components: an atmospheric model, a stellar contamination model, and a Gaussian process. We describe each of these in the following subsections.

\subsection{Atmospheric Model} \label{subsec:atmosphere_setup}

We assume free chemistry in our atmospheric model and, motivated by giant exoplanet \textit{JWST} retrievals in the literature \citep{tsai2023, fu2024, crossfield2025} and equilibrium abundance predictions at TOI-3235~b's temperature ($T_{\rm eq} \sim 600$~K) \citep{zahnle2009, madhusudhan2012, heng2016}, we include opacity contributions from the following molecular species: \ce{CO2} \citep{yurchenko2020exomol}, \ce{CH4} \citep{yurchenko2024exomol}, \ce{H2O} \citep{polyansky2018exomol}, \ce{NH3} \citep{coles2019exomol}, \ce{HCN} \citep{barber2014exomol}, \ce{CO} \citep{li2015rovibrational}, \ce{SO2} \citep{underwood2016exomol}, \ce{H2S} \citep{azzam2016exomol}, \ce{C2H2} \citep{chubb2020exomol}, and \ce{C2H4} \citep{gordon2022hitran}. These species exhibit strong or potentially overlapping spectral features across the NIRSpec/PRISM wavelength range (0.6--5.6~\micron), making them relevant contributors to the observed transmission spectrum. We compute forward models on a wavelength grid spanning 0.5--5.7~\micron\ at a constant resolving power of $R = 20{,}000$. The mixing ratio of each molecular species is treated as a free parameter, and we assume a bulk composition of \ce{H2} and \ce{He}, with the \ce{He} abundance fixed to its solar value relative to \ce{H2}, such that $\mathrm{He}/\mathrm{H}_2 = 0.17$. 

The model atmospheres span pressures from $10^{-7}$ to $10^{3}$~bar, divided into $100$ log-uniformly spaced layers. We retrieve the reference pressure $P_{\rm ref}$, fixing the reference radius to the planet's white-light radius of $11.4\,R_\oplus$ \citep{hobson2023}. It is also possible to fit the planetary radius at a fixed reference pressure, but the choice of the reference parameter has a negligible impact on the retrieval results \citep{welbanks2019degeneracies}. We set the planetary mass to $211\,M_\oplus$, which POSEIDON uses to calculate the planetary surface gravity corresponding to the observed radius, and the stellar radius to $0.3697\,R_\odot$ \citep{hobson2023}.

We include a gray cloud deck parameterized by the log of the cloud top pressure, $\log P_{\rm cloud}$, and a power-law haze slope parameterized by a haze amplitude $\log a$ and a haze scattering slope $\gamma$. We do not consider patchy clouds and assume cloud and haze uniformity around the terminator. We also retrieve a white noise amplitude term $\sigma_{\rm w}$ to account for underestimated error bars and combine it with the reported measurement uncertainties:
\begin{equation}\label{eq:sigma_eff}
    \sigma_{\rm eff}^2(\lambda)
    = \sigma_{\delta}^2(\lambda) + \sigma_w^2,
\end{equation}
where $\sigma_{\delta}(\lambda)$ is the measured uncertainty on the transit depth.

We include a flux offset parameter to account for the unknown true transit depth that may be blanketed by effects such as stellar contamination. We emphasize that, although the flux offset is partially degenerate with the reference pressure $P_{\rm ref}$, they represent distinct physical effects. $P_{\rm ref}$ sets the pressure level of the planetary radius where we begin integrating the atmosphere, whereas the flux offset applies a shift to the transit depths to account for uncertainty. We therefore include both parameters in our retrievals. A summary of our priors for our atmospheric retrieval is presented in Table \ref{tab:priors}.

\subsection{Stellar Contamination Model} \label{subsec:SC_setup}

For the stellar contamination model, we adopt $T_{\rm eff} = 3388.8$~K and $\log g_* = 4.8976$ for the stellar parameters \citep{hobson2023} and interpolate solar-metallicity \texttt{BT-SETTL} synthetic spectra using \texttt{scipy.interpolate.RegularGridInterpolator}. We then rebin the interpolated spectra onto our model wavelength grid with \texttt{pysynphot} \citep{lim2015pysynphot}. The underlying BT-SETTL models are from the CIFIST grid of \citet{allard2013bt}.

We adopt the \texttt{BT-SETTL} grid as a representative set of stellar synthetic spectra for the contamination model because it spans effective temperatures from 400 to 70,000~K \citep{allard2013bt}, encompassing the cool temperatures relevant for M dwarfs and their stellar heterogeneities. Other stellar atmosphere grids may produce different spectral shapes and therefore represent a source of systematic uncertainty in the inferred contamination spectrum. A full comparison among stellar spectral grids is beyond the scope of this current work.

We fit for the photosphere temperature and consider both a one-component and two-component contamination model, following the formalism of \citet{rackham2018}. In the one-component case, we retrieve a single heterogeneity temperature and covering fraction $f_{\rm spot}$ \citep[e.g.][]{bennett2025additional,rathcke2021hst}; in the two-component case, we retrieve two temperatures and covering fractions $(f_{\rm spot}, f_{\rm fac})$ to represent, e.g., spots and faculae \citep[e.g.][]{fournier2024near}. We choose a log-uniform prior on the heterogeneity covering fraction between $10^{-3}$ and 1, which allows the retrieval to explore several orders of magnitude without favoring large covering fractions.

Following \citet{rackham2018}, we model stellar contamination as a wavelength-dependent multiplicative factor $\epsilon(\lambda)$ applied to the planetary transmission spectrum,
\begin{equation}\label{eq:epsilon}
    \delta_{\rm obs}(\lambda) = \delta_{\rm atm}(\lambda)\,\epsilon(\lambda),
\end{equation}
where $\delta_{\rm obs}(\lambda)$ is the observed transit depth and $\delta_{\rm atm}(\lambda)$ is the intrinsic atmospheric transit depth from \texttt{POSEIDON}. We do not observe any spot-crossing events in our light curve, so we assume that the transit chord covers a homogeneous photosphere. In the one-component case then, the occulted intensity is $S_{\rm phot}(\lambda)$ and the disk-averaged stellar intensity is

\begin{equation}
    S_{\rm disk}(\lambda) = \left(1 - f_{\rm spot}\right) S_{\rm phot}(\lambda) + f_{\rm spot}\,S_{\rm spot}(\lambda),
\end{equation}

where $f_{\rm spot}$ is the spot covering fraction and $S_{\rm phot}(\lambda)$ and $S_{\rm spot}(\lambda)$ are the photosphere and spot spectra.

Extended to the two-component model, the disk-averaged stellar intensity becomes
\begin{equation}
\begin{aligned}
    S_{\rm disk}(\lambda) &= \left(1 - f_{\rm spot} - f_{\rm fac}\right) S_{\rm phot}(\lambda) \\
    &\quad + f_{\rm spot}\,S_{\rm spot}(\lambda)
    + f_{\rm fac}\,S_{\rm fac}(\lambda).
\end{aligned}
\end{equation}
where $f_{\rm fac}$ is the facular covering fraction and $S_{\rm fac}(\lambda)$ is the facula spectrum.

The stellar contamination factor is then

\begin{equation}
    \epsilon(\lambda) = \frac{S_{\rm phot}(\lambda)}{S_{\rm disk}(\lambda)}.
\end{equation}

which is applied as a multiplicative factor as shown in Equation~\ref{eq:epsilon}.

\subsection{Gaussian Process} \label{subsec:GP_setup}
To model stellar contamination signals that might not be captured by the model described in the previous section, or to model residual correlated instrumental and/or astrophysical noise, we include a multiplicative GP that fits the residuals from the atmospheric and stellar contamination model, following Equation 1 in \citealt{espinoza2025dreams}. We use a Mat\'ern~3/2 kernel implemented with \texttt{george} \citep{george}, which introduces two additional free hyperparameters: the length scale $\rho$ and the amplitude $\sigma_{\rm GP}$. The white noise amplitude $\sigma_{\rm w}$ is included in the diagonal of the covariance matrix. Both Mat\'ern and squared-exponential kernels are commonly used in exoplanet retrievals, and we run retrievals with each kernel, finding that they yield similar results. Thus, we use the Mat\'ern~3/2 kernel for all our GP implementations due to its flexibility in capturing sharp features while modeling correlated noise.

\subsection{Model Comparison} \label{subsec:modelcomp}

We choose to use Bayesian evidences as our metric for model comparison, as it automatically incorporates a penalty for model complexity from integrating over the prior space, making it an ideal choice for balancing model fit and complexity within a Bayesian framework.

% The Bayes Factor quantifies the relative evidence between two competing models, and is defined as the ratio of their Bayesian evidences \( Z \):
% \[
% B_{10} = \frac{\mathrm{P(Data \mid Model~1)}}{\mathrm{P(Data \mid Model~0)}} = \frac{Z_1}{Z_0}.
% \]
% In its logarithmic form, the Bayes Factor can be expressed as:
% \[
% \log B_{10} = \log(Z_1) - \log(Z_0) = \Delta \log Z.
% \]

While the translation of Bayesian evidence to frequentist measures is occasionally done due to higher familiarity with sigma thresholds, we avoid this practice because such conversions can overstate the significance of results \citep{kipping2025exoplaneteers}. We also include reduced chi-squared values ($\chi^2_\nu$), calculated as
\begin{equation}\label{eq:redchisquared}
    \chi^2_\nu = \frac{1}{N-k}
    \sum_{i=1}^{N}
    \left(\frac{\delta_i - m_i}{\sigma_i}\right)^2,
\end{equation}
where $N$ is the number of spectral bins, $k$ is the number of free parameters, $\delta_i$ is the measured transit depth in the $i$th spectral bin, $m_i$ is the corresponding model transit depth, and $\sigma_i$ is the effective uncertainty in the $i$th spectral bin ($\sigma_{\rm eff}(\lambda)$ from Equation~\ref{eq:sigma_eff}, using the median retrieved $\sigma_w$). For retrievals including a GP, this quantity serves only as a residual diagnostic, since the full likelihood also includes the GP covariance.

\subsection{Derived Atmospheric Metallicity and C/O Ratio}

We compute the atmospheric metallicity and C/O ratio from the retrieved free molecular abundances. Because the individual abundances are assigned independent log-uniform priors, the implied priors on these derived quantities are not uniform. For metallicity, the prior volume preferentially weights high values because $Z$ depends on the sum of linear heavy-element abundances. The C/O ratio is similarly affected because it is computed from ratios of sums over carbon- and oxygen-bearing species.

We correct for these induced prior-volume effects using importance reweighting. For metallicity, each posterior sample is weighted by the inverse of the induced prior density in $\log_{10}(Z/Z_\odot)$, equivalent to adopting a target prior uniform in $\log_{10}(Z/Z_\odot)$. For C/O, we apply the analogous correction in $\log_{10}({\rm C/O})$, corresponding to a target prior uniform in $\log_{10}({\rm C/O})$. We choose a log uniform target prior for C/O because the free-chemistry abundance priors allow C/O to span several orders of magnitude, particularly when the oxygen-bearing abundances are small.

The resulting metallicity and C/O posteriors should therefore be interpreted as prior-corrected derived quantities from the free-chemistry retrievals, not as quantities retrieved directly under chemical equilibrium. We describe the reweighting procedure and induced priors in Appendix~\ref{app:D}.

\section{Results} \label{sec:results}

\begin{figure*}
    \centering
    \includegraphics[width=0.7\linewidth]{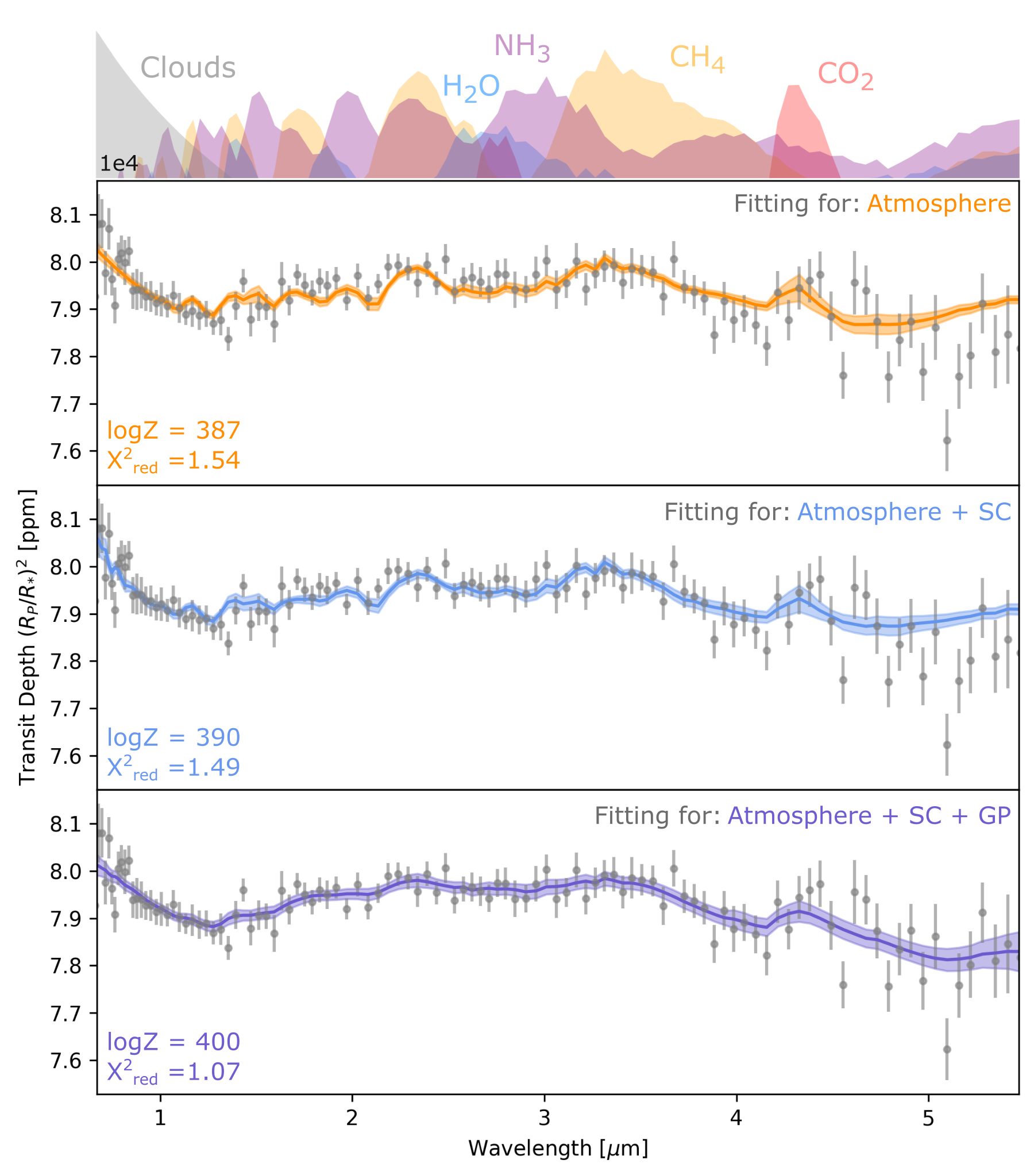}
    \caption{Retrieved spectra from each retrieval. The solid lines represent the median spectrum, and the shaded regions show the 1$\sigma$ bounds. The first panel shows the retrieval only fitting for the atmospheric model, the second shows the retrieval fitting for the atmospheric model and the stellar contamination model, and the third shows the retrieval fitting for the atmospheric model, the stellar contamination model, and a Gaussian process. The top region shows the spectral contribution plots corresponding to the first retrieval only fitting for the atmospheric model.}
    \label{fig:all_retrievals}
\end{figure*}

\begin{figure*}
    \centering
    \includegraphics[width=\linewidth]{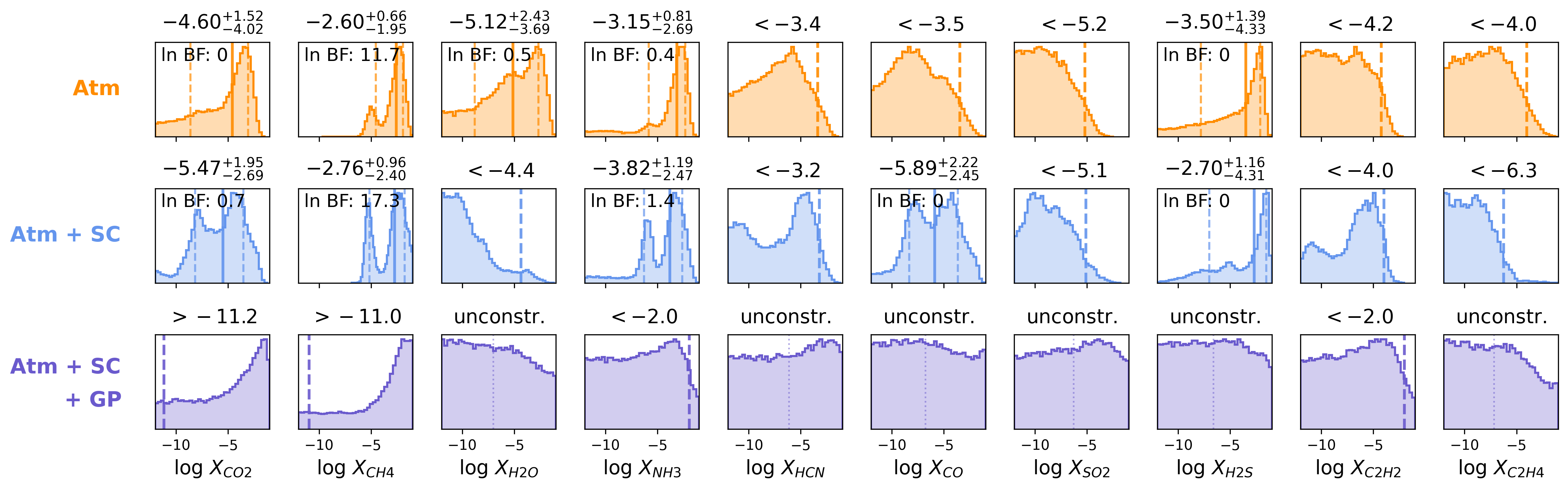}
    \caption{Retrieved trace abundances for all three retrievals. Each panel shows the posterior distribution of $\log X_i$ for one molecular species. Numerical labels above each panel summarize the abundance constraint: for molecules with two-sided bounded posteriors, we quote the median and 1$\sigma$ bounds; for molecules with only one bound, we quote either the 5th-percentile lower limit or the 95th-percentile upper limit; and for posteriors that reach both prior edges, we label the abundance as unconstrained. For molecules with two-sided bounds, we include the Bayes factor.}
    \label{fig:all_trace}
\end{figure*}

We present our results beginning with a retrieval only fitting for the atmospheric model and hierarchically incorporating first the stellar contamination model and then the GP. The retrieved spectra from the retrievals are shown in Figure~\ref{fig:all_retrievals}, the retrieved abundances are shown in Figure~\ref{fig:all_trace}, and the stellar contamination and haze parameters are shown in Figure~\ref{fig:stellar_parameters}. We provide chemical equilibrium abundance predictions in Appendix~\ref{app:A} for comparison.

\begin{figure*}
\centering
\includegraphics[width=0.7\linewidth]{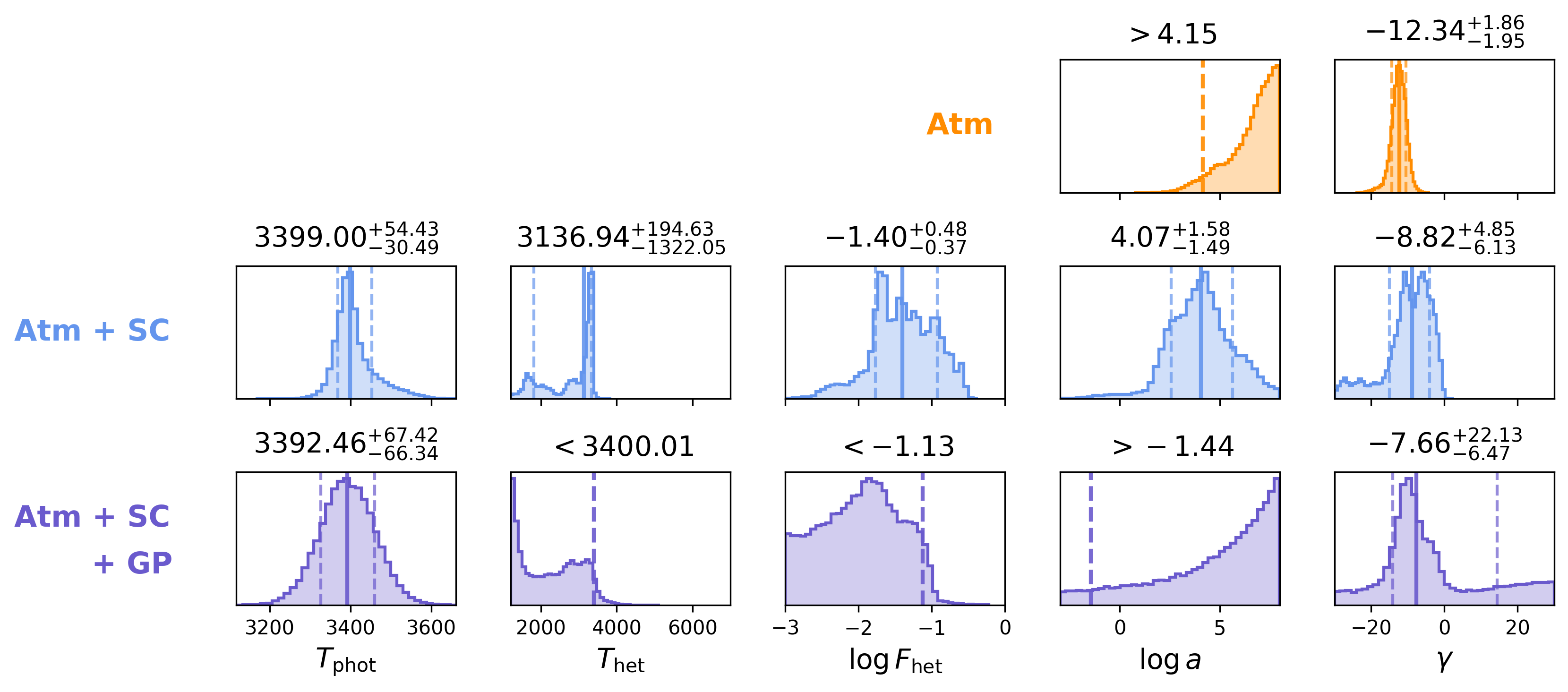}
\caption{
Posterior distributions for the stellar contamination and haze parameters across the three retrieval models. From left to right: photosphere temperature ($T_{\rm phot}$), heterogeneity (spot) temperature ($T_{\rm het}$), log heterogeneity (spot) covering fraction ($\log f_{\rm spot}$), log haze scattering amplitude ($\log a$), and haze scattering slope ($\gamma$). Numerical labels above each panel summarize the parameter constraint: for parameters with two-sided bounded posteriors, we quote the median and 1$\sigma$ bounds; for parameters with only one bound, we quote either the 5th-percentile lower limit or the 95th-percentile upper limit.
}
\label{fig:stellar_parameters}
\end{figure*}

\subsection{Atmospheric Model Only} \label{subsec:atmosphere_results}

We first run a retrieval only fitting for the atmospheric model, as described in Section~\ref{subsec:atmosphere_setup}. The retrieved spectra are shown in the first panel of Figure~\ref{fig:all_retrievals}, the retrieved trace abundances are shown in the first row of Figure~\ref{fig:all_trace}, and haze parameters are shown in Figure~\ref{fig:stellar_parameters}. The corner plot of all retrieved posteriors is shown in Appendix~\ref{app:B}, Figure~\ref{fig:atm_cp}. We obtain a Bayesian evidence value of $\ln Z = 387$ and a reduced chi-squared value of $\chi^2_\nu = 1.54$.

We find strong evidence for hazes that account for the observed slope at short wavelengths. The haze scattering slope and amplitude converge to $\gamma = -12.34^{+1.86}_{-1.95}$ and $\log a = 6.90^{+0.78}_{-1.60}$, with the haze amplitude saturating at the upper edge of our already broad prior ($\log a \in [-3, 8]$).

Haze formation is especially efficient in cool atmospheres at equilibrium temperatures $T_{\rm eq} < 1000$\,K \citep{kawashima2019, gao2021}, leading to an increased opacity at bluer wavelengths that causes transit depth to increase \citep{sing2016, sedaghati2017, may2019}. Rayleigh scattering by small aerosol particles creates a wavelength dependence with a scattering slope of $-4$ \citep{des2008rayleigh}, although steeper slopes have been observed for hot Jupiters due to enhanced scattering effects \citep{ohno2020}. 

However, cool star spots on the host star can imprint similar slope-like features in the transmission spectrum, as spots suppress blue light \citep{mccullough2014water, espinoza2019}. In this atmosphere-only retrieval, any such stellar contamination must be absorbed by the atmospheric parameters, most notably the haze slope and amplitude. We therefore interpret the extreme haze solution not as a robust physical measurement, but as evidence that the observed slope captured only by the haze parameterization is likely a combination of both hazes and stellar contamination, motivating an additional stellar contamination component into our model. 

The model also shows comparatively larger residuals at wavelengths beyond $4.7~\mu$m. To quantify whether these long-wavelength residuals correspond to a systematic offset, we compute the uncertainty-weighted mean residual over the 13 spectral bins at $\lambda > 4.7~\mu$m. For each bin, we define the residual as $r_i = \delta_i - m_i$, where $\delta_i$ is the measured transit depth and $m_i$ is the median retrieved model transit depth. We combine the data and model uncertainties $\sigma_{r,i}^2 = \sigma_{\delta,i}^2 + \sigma_{m,i}^2$, and compute the weighted mean residual $\bar{r}$ and its uncertainty. For the atmosphere-only retrieval, we find $\bar{r} = -752 \pm 191$ ppm, which corresponds to a $3.9\sigma$ offset from zero, suggesting that the atmosphere-only model does not fully capture the structure present beyond $4.7~\mu$m.

\begin{figure*}
    \centering
   \includegraphics[width=0.6\linewidth]{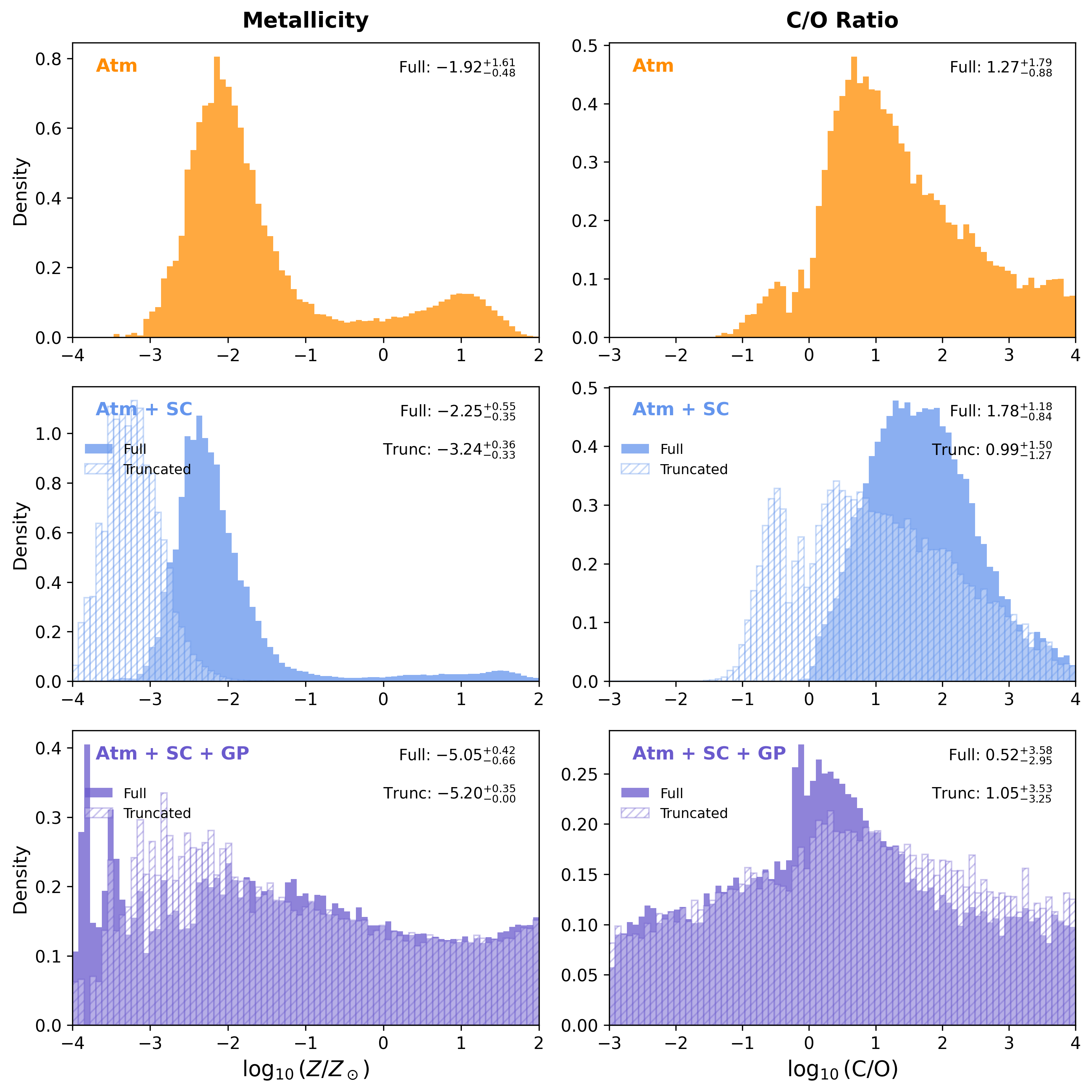}
    \caption{Prior-corrected metallicity and C/O posteriors for the full-spectrum retrievals, with truncated-spectrum atm+SC and atm+SC+GP results shown for comparison.}
    \label{fig:metallicity}
\end{figure*}

For the purpose of model comparison, we summarize the molecular abundance constraints obtained in this atmosphere-only retrieval, shown in Figure~\ref{fig:all_trace}. Among the ten trace species considered, methane is the only constrained detection, with $\log X_{\mathrm{CH_4}} = -2.60^{+0.66}_{-1.95}$ (ln BF = 11.7). However, several other species have peaks in their posteriors, suggesting potential evidence for their presence: $\mathrm{CO_2}$, $\mathrm{H_2O}$, $\mathrm{NH_3}$, and $\mathrm{H_2S}$. The posteriors exhibit a bimodality, most visible in the $\mathrm{CH_4}$ posterior, which propagates into the metallicity posterior, shown in Figure~\ref{fig:metallicity}. This arises from a degeneracy between atmospheric metallicity and haze opacity, where either lower abundances with weaker hazes or higher abundances with stronger hazes can reproduce the observed spectrum (see the $\log a$ vs. CH$_4$ correlation in Figure~\ref{fig:atm_cp}). Nonetheless, the prior-corrected metallicity strongly prefers sub-solar values with $[\mathrm{M/H}] = -1.93^{+1.60}_{-0.48}$. The CH$_4$ posterior exhibits two modes: a super-solar mode and a sub-solar mode (Figure~\ref{fig:all_trace}). However, after applying the prior correction described in Appendix~\ref{app:D}, the super-solar mode is downweighted while the sub-solar mode is upweighted, resulting in the sub-solar metallicity preference seen in Figure~\ref{fig:metallicity}.

We infer a prior-corrected C/O ratio of $\log(\mathrm{C/O}) = 1.23^{+1.61}_{-0.85}$ (${\rm C/O} \sim 17^{+675}_{-15}$), which prefers a super-solar value. However, the broad distribution reflects the fact that most individual carbon- and oxygen-bearing species remain weakly constrained.

We stress, however, that these results are obtained under the assumption that all wavelength-dependent structure in the spectrum is of planetary origin, which is unlikely due to the presence of stellar contamination. Therefore, the atmosphere-only retrieval is likely biased, with stellar signals being interpreted as atmospheric hazes and molecular features. For this reason, we do not adopt these results as our science results and instead treat this retrieval as a diagnostic baseline and motivation to introduce additional model components to capture stellar effects.

\subsection{Atmospheric Model + Stellar Contamination Model} \label{subsec:SC_results}

We run a second retrieval fitting for the atmospheric model and a stellar contamination model as described in Section~\ref{subsec:SC_setup}. We run retrievals fitting for both the one-component and two-component stellar contamination model. In both cases, the heterogeneity temperatures are assigned uniform priors between 1200 and 7000~K, and the corresponding covering fractions are assigned log-uniform priors between $10^{-3}$ and 1 (see Table~\ref{tab:priors}). For the two-component model, we additionally enforce that one heterogeneous component is hotter than the other to break the degeneracy between the two.

We find that the Bayesian evidence is higher for the one-component stellar contamination model with $\Delta \ln Z = 5$. We therefore use the one-component stellar contamination model for all retrievals. The retrieved spectra are shown in the second panel of Figure~\ref{fig:all_retrievals}, the retrieved trace abundances are shown in the second row of Figure~\ref{fig:all_trace}, and haze and stellar contamination parameters are shown in Figure~\ref{fig:stellar_parameters}. The corner plot of all retrieved posteriors is shown in Appendix~\ref{app:B}, Figure~\ref{fig:atm_sc_cp}. Once we add the stellar contamination model to the atmospheric model, the Bayesian evidence increases by $\Delta \ln Z = 3$ relative to the atmosphere-only case. The reduced chi-squared also improves from $\chi^2_\nu = 1.54$ to $\chi^2_\nu = 1.49$.

The inferred haze properties now converge to $\gamma = -8.82^{+4.15}_{-4.83}$ and $\log a = 4.07^{+1.58}_{-1.48},$ indicating weaker hazes than in the atmosphere-only retrieval ($\gamma = -12.34^{+1.86}_{-1.95}$, $\log a = 6.90^{+0.78}_{-1.60}$). This behavior is consistent with the initial short-wavelength slope being jointly fit by both the haze parameterization and the stellar contamination. We retrieve a heterogeneity temperature of $T_{het} = 3140^{+190}_{-1320}$ K and a heterogeneity covering fraction of $\log f_{het} = -1.40^{+0.48}_{-0.37}$, which corresponds to around 2 to 12 percent.

As in the atmosphere-only case, methane remains the only constrained molecular detection (ln BF = 17.3). The CH$_4$ posterior exhibits two modes: a lower-abundance mode at $\log X_{\mathrm{CH_4}} = -5.18^{+0.46}_{-0.34}$ and a higher-abundance mode at $\log X_{\mathrm{CH_4}} = -2.33^{+0.72}_{-0.69}$. Several other species again have peaks in their posteriors, suggesting potential evidence for their presence: CO$_2$, NH$_3$, CO, and H$_2$S. Relative to the atmosphere-only retrieval, the H$_2$O posterior shifts to an upper limit, without the tentative peak seen in the atmosphere-only model. This illustrates the impact of the stellar contamination model: the optical slope can be attributed to cool star spots, which account for trace amounts of water. Given that H$_2$O is expected both in cool giant-planet atmospheres and in cool star spots, neglecting stellar contamination can bias water inferences toward artificially high abundances. We again observe bimodality in the posteriors, driven by degeneracies among haze opacity, molecular abundances, and stellar contamination. The stellar contamination model adds another way to reproduce the short-wavelength slope, allowing the observed spectrum to be fit by different combinations of cool star spots, haze opacity, and molecular abundance.

In this atm+SC retrieval, we find a strong preference for sub-solar prior-corrected metallicity, with $[\mathrm{M/H}] = -2.25^{+0.55}_{-0.35}.$ The prior-corrected C/O ratio is shifted toward super-solar values, with $\log(\mathrm{C/O}) = 1.78^{+1.18}_{-0.84}$ (${\rm C/O} \sim 60^{+850}_{-50}$), though the posterior remains broad.

Although the atm+SC model provides a modestly better statistical fit than the atmosphere-only model ($\Delta \ln Z = 3$; $\chi^2_\nu$ improves from 1.54 to 1.49), it remains limited by the simplicity of the stellar contamination prescription. In particular, the long-wavelength residuals remain comparable to those from the atmosphere-only retrieval. Using the same uncertainty-weighted mean residual metric over the 13 spectral bins at $\lambda > 4.7~\mu$m, we find $\bar{r} = -740 \pm 191$ ppm, corresponding to a $3.9\sigma$ offset from zero. This is nearly identical to the atmosphere-only result, indicating that the addition of the one-component stellar contamination model does not substantially reduce the systematic long-wavelength offset.

To test whether this long-wavelength structure influences the retrieved atmospheric properties, we repeat the atm+SC retrieval after truncating the transmission spectrum at $4.7~\mu$m. For the truncated atm+SC retrieval, we obtain $\ln Z = 362.31$ and $\chi^2_\nu = 1.20$. We caution that the Bayesian evidence for the truncated retrieval should not be directly compared to the evidence from the full-spectrum retrieval because the truncated and full spectra contain different numbers of data points covering different wavelength ranges. Instead, we use the truncated retrieval as a sensitivity test to assess how strongly the inferred atmospheric properties depend on the long-wavelength data.

The truncated and untruncated atm+SC retrievals yield different posterior distributions, as shown in Appendix~\ref{app:C}. When the data beyond $4.7~\mu$m are removed, the CO$_2$ abundance decreases, the H$_2$O abundance increases, suggesting a tentative water detection that is not present in the full-wavelength retrieval. The molecular abundance posteriors also become largely unimodal rather than bimodal. This result may be more physically plausible, as CO$_2$ is not expected to be abundant at the equilibrium temperature of TOI-3235~b under standard chemical equilibrium conditions, as shown in Figure~\ref{fig:vulcan_eq}. The truncated retrieval also shifts the prior-corrected C/O to lower values from $\log({\rm C/O}) = 1.78^{+1.18}_{-0.84}$ (${\rm C/O} \sim 60^{+850}_{-50}$) to $\log({\rm C/O}) = 0.99^{+1.50}_{-1.27}$ (${\rm C/O} \sim 10^{+300}_{-9}$), though the posterior remains broad, as shown in Figure~\ref{fig:metallicity}. The prior-corrected metallicity shifts to even lower values, with $[\mathrm{M/H}] = -3.24^{+0.36}_{-0.33}$.

Removing the long-wavelength region changes the inferred molecular abundances, metallicity, and C/O ratio, demonstrating that the deterministic atm+SC interpretation is not only model-dependent but also sensitive to the wavelength range included in the retrieval. When the model is restricted to deterministic atmospheric and stellar-contamination components, residual structure at long wavelengths can only be absorbed into the retrieved atmospheric parameters, producing inferences that are likely biased by model inadequacy. This provides further motivation for including a GP component to flexibly marginalize over residual wavelength-correlated structure.

\subsection{Atmospheric Model + Stellar Contamination Model + Gaussian Process} \label{subsec:GP_results}
To account for correlated noise that the atmospheric model and stellar contamination model fail to fit, we run a third retrieval fitting for the atmospheric model, the stellar contamination model, and a Gaussian process, as described in Section~\ref{subsec:GP_setup}. The retrieved spectra are shown in the third panel of Figure~\ref{fig:all_retrievals}, the retrieved trace abundances are shown in the third row of Figure~\ref{fig:all_trace}, and haze and stellar contamination parameters are shown in Figure~\ref{fig:stellar_parameters}. The corner plot of all retrieved posteriors is shown in Appendix~\ref{app:B}, Figure~\ref{fig:atm_sc_gp_cp}. The Bayesian evidence increases by $\Delta \ln Z = 10$ relative to the atm+SC retrieval and $\Delta \ln Z = 13$ relative to the atmosphere-only retrieval. The reduced chi-squared also improves by $\Delta \chi^2_\nu = -0.42$ and $\Delta \chi^2_\nu = -0.47$, respectively. The improvement is especially apparent at long wavelengths; using the same uncertainty-weighted mean residual metric over the 13 spectral bins at $\lambda > 4.7~\mu$m, we find $\bar{r} = -187 \pm 196$ ppm, corresponding to a $1.0\sigma$ offset from zero, an improvement from the $3.9\sigma$ offset from both the atmosphere-only and atm+SC retrievals. This indicates that the additional GP component provides a better fit to the data, particularly in the region beyond $4.7~\mu$m.

To demonstrate that the long-wavelength structure no longer drives the retrieval results once the GP is included, we repeat the atm+SC+GP retrieval on the spectrum truncated at $4.7~\mu$m. For the truncated atm+SC+GP retrieval, we obtain $\ln Z = 362.04$ and $\chi^2_\nu = 1.04$. We once again caution that this evidence should not be directly compared to the full-spectrum evidence because the truncated and full spectra contain different numbers of data points covering different wavelength ranges.

Unlike the atm+SC case, the truncated and untruncated atm+SC+GP retrievals yield consistent abundance constraints, shown in Appendix~\ref{app:C}. The metallicity and C/O posteriors are also broadly consistent between the two atm+SC+GP retrievals. This indicates that the GP effectively marginalizes over the uncertainty associated with the long-wavelength correlated structure, preventing it from biasing the inferred atmospheric parameters.

The cost of this marginalization is that the abundance constraints broaden. Rather than forcing the physical atmospheric model to explain residual structure it cannot capture, the GP propagates this uncertainty into the posterior distributions. As a result, the inferred prior-corrected metallicity and C/O ratio are both largely uninformative, as shown in Figure~\ref{fig:metallicity}.

Although the abundance constraints are broad, substantial posterior mass remains at elevated CO$_2$ and CH$_4$ abundances, as seen in Figure~\ref{fig:all_trace}. The CO$_2$ elevated abundance appears once the GP absorbs the broader long-wavelength residual structure, allowing the atmospheric model to fit localized structure near the $4.3~\mu$m CO$_2$ opacity feature. The CH$_4$ elevated abundance is similar to that in the retrievals without a GP and is driven by spectral structure near the $\sim2.3$ and $3.3~\mu$m CH$_4$ bands. We therefore do not interpret these abundance preferences as robust detections but rather as tentative features that remain sensitive to the flexibility of the GP model.

With the addition of a GP, the haze scattering slope becomes shallower, converging to $\gamma = -7.66^{+22.13}_{-6.47}$ because both the deterministic stellar contamination model and the GP can now account for the steep optical slope, reducing the haze contribution required to fit the data. The spot temperature shifts from a constrained value of $T_{\rm spot} = 3137^{+195}_{-1322}$~K in the atm+SC retrieval to an upper limit in the atm+SC+GP retrieval. However, the photosphere temperature remains well-constrained at $T_{\rm phot} = 3392^{+67}_{-66}$~K, consistent with the atm+SC value of $T_{\rm phot} = 3399^{+54}_{-30}$~K. This demonstrates that including the GP selectively broadens parameters that are degenerate with correlated spectral structure, while parameters constrained by independent spectral information retain meaningful constraints.

We further test whether the inclusion of a GP can cause the retrieval to erase atmospheric and stellar features that are physically present in the data. To do this, we generate a synthetic transmission spectrum using the median parameters from the atm+SC retrieval, inject observational noise, and then retrieve this simulated spectrum with the more flexible atm+SC+GP model. In this controlled case, where the deterministic atmospheric and stellar contamination model is sufficient to describe the data, the GP does not dominate the fit: the injected atmospheric and stellar parameters are recovered within the posterior uncertainties. This experiment indicates that the broad posteriors obtained for the real TOI-3235~b spectrum are not simply an unavoidable consequence of including a GP. Instead, they likely reflect the presence of wavelength-correlated structure in the real data that is not fully captured by the deterministic atmospheric and stellar contamination models. We provide the details of this injection-retrieval test in Appendix~\ref{app:gp_injection}.

\section{Discussion} \label{sec:discussion}

\subsection{Model dependence of the transmission spectrum interpretation}
\label{sec:discussion_model_dependence}

We find that the inferred atmosphere of TOI-3235~b depends on how stellar contamination is modeled. In the atmosphere-only retrieval, all wavelength-dependent structure in the transmission spectrum must be explained by planetary opacity sources. This model therefore attributes the observed spectral features to a combination of atmospheric hazes, clouds, and molecular abundances. However, stellar heterogeneities can imprint wavelength-dependent signals that mimic or distort planetary absorption features. Ignoring this effect can therefore bias the retrieved atmospheric properties.

Including a parametric stellar contamination model changes the inferred atmospheric interpretation. In the atm+SC retrieval, part of the observed spectral structure is assigned to unocculted stellar heterogeneities rather than to the planet alone. This reduces some of the biases present in the atmosphere-only fit, including the tendency to interpret stellar H$_2$O features as planetary H$_2$O absorption. The resulting atm+SC solution is physically interpretable and we discuss its physical implications in Section~\ref{sec:discussion_atm_sc}. Nevertheless, it remains dependent on the assumed stellar contamination prescription and on the wavelength range included in the retrieval.

The atm+SC+GP retrieval provides the most conservative interpretation of the current transmission spectrum. In this model, the GP absorbs residual wavelength-correlated structure that is not captured by the deterministic atmosphere+SC model. As a result, many atmospheric parameters become substantially less constrained. We do not interpret this as evidence that the planet necessarily lacks atmospheric features; rather, it indicates that the current data do not contain enough information within our retrieval framework to distinguish between planetary absorption, stellar contamination, and residual wavelength-correlated structure. These retrievals demonstrate that stellar contamination is a dominant limitation in interpreting the transmission spectrum of TOI-3235~b, and atmospheric inferences are therefore model-dependent.

\subsection{Conditional atmospheric interpretation of the atm+SC retrieval}
\label{sec:discussion_atm_sc}

Although the atm+SC+GP retrieval provides the most conservative interpretation, the deterministic atm+SC retrieval remains useful as a conditional physical scenario. The atm+SC model yields a CH$_4$ posterior with significant probability away from the prior bounds, and posterior peaks for CO$_2$, NH$_3$, CO, and H$_2$S, though these species are not robustly detected. The retrieved abundances correspond to a sub-solar atmospheric metallicity and a potentially supersolar but uncertain C/O ratio. The truncated atm+SC retrieval, which excludes data beyond 4.7~$\mu$m, points to a well-constrained CH$_4$ abundance and posterior peaks for H$_2$O and C$_2$H$_2$, though these species are not robustly detected. The retrieval also shows a lower CO$_2$ abundance and a shift toward lower C/O values, though uncertainties remain broad.

In the core-accretion paradigm, giant-planet atmospheres are generally expected to be enriched relative to their host stars because of the accretion or dissolution of solids during formation \citep{mordasini:2016,mordasini2024}. This expectation is especially relevant for GEMS, whose formation around low-mass stars may require enhanced solid surface densities. While gravitational instability can produce planets with compositions closer to the disk gas \citep{kratter2010,boss2023}, it does not predict atmospheres that are strongly depleted in heavy elements. Thus, if TOI-3235~b truly has an atmospheric metallicity as low as suggested by the atm+SC retrieval on both the full and truncated data, it would deviate from standard expectations of core accretion or gravitational instability and might indicate accretion of solid-depleted gas. Depletion of solids via radial drift around M dwarfs might be more efficient due to shorter radial drift timescales as evidenced by the steepening of the disk dust mass to stellar mass relation \citep{pascucci2016}.

The high C/O ratio should be interpreted with caution. The prior-corrected value, $\log({\rm C/O}) = 1.78^{+1.18}_{-0.84}$ (${\rm C/O} \sim 60^{+850}_{-50}$) for the full spectrum and $\log({\rm C/O}) = 0.99^{+1.50}_{-1.27}$ (${\rm C/O} \sim 10^{+300}_{-9}$) for the truncated spectrum, extends over orders of magnitude. This retrieved C/O reflects the ratio of carbon- to oxygen-bearing molecules included in the retrieval, not necessarily the bulk elemental C/O of the planet. In such a retrieval framework, when certain species are poorly constrained---for instance, if an O-bearing species is weakly detected while a C-bearing species has a broad prior---the inferred C/O can reach unphysically high values that do not reflect the true elemental composition. Such extreme bulk C/O ratios are difficult to produce via standard planet formation pathways. The accretion of oxygen-depleted gas outside the H$_2$O snowline can raise the atmospheric C/O to order unity, while substantial planetesimal enrichment generally drives the C/O ratio back toward stellar or sub-stellar values \citep{oberg2011}. The extremely high C/O values from the atm+SC retrieval are therefore unlikely to represent a robust compositional or formation signature. Therefore, we do not use retrieved C/O values to draw formation implications and interpret them only as highlighting unconstrained molecular abundances.

\begin{figure*}
    \centering
    \includegraphics[width=0.8\linewidth]{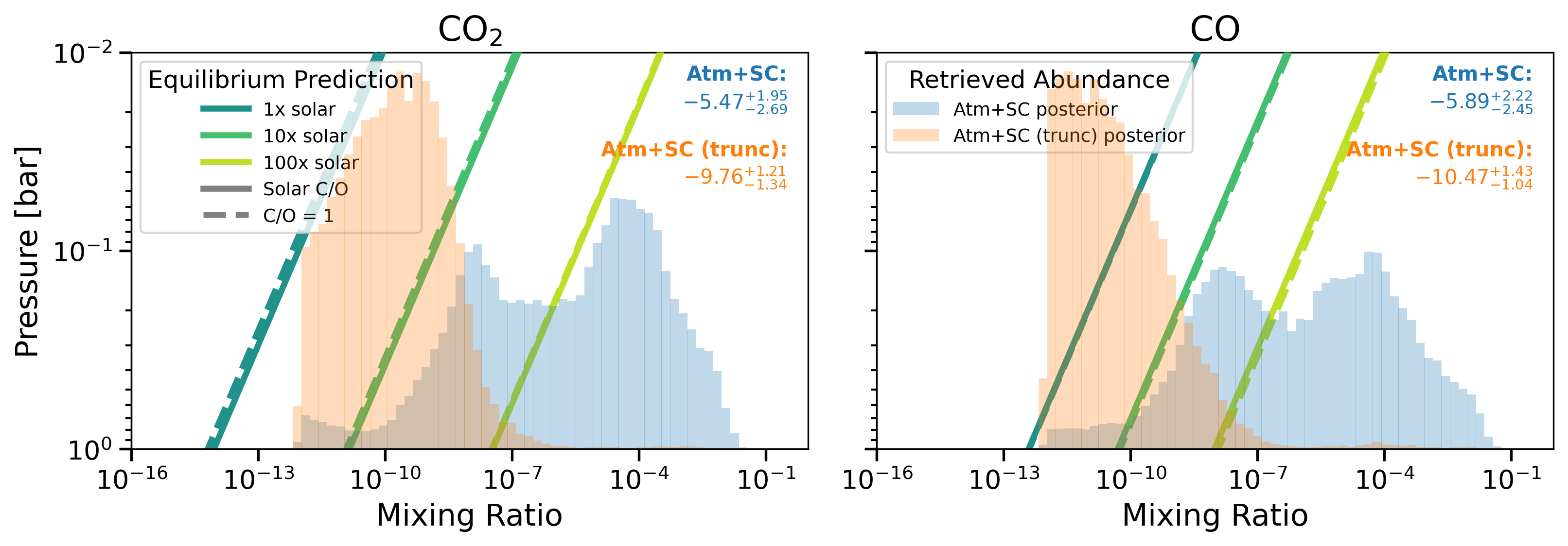}
    \caption{
    Comparison between the atm+SC abundance posteriors and VULCAN chemical equilibrium predictions for CO$_2$ and CO. The blue histograms show the full-spectrum atm+SC posterior distributions, while the orange histograms show the truncated atm+SC posterior distributions (excluding data beyond 4.7~$\mu$m). The black solid and dashed vertical lines mark the posterior median and 1$\sigma$ bounds. Colored curves show VULCAN equilibrium abundance profiles for 1, 10, and 100$\times$ solar metallicity, with solid lines indicating solar C/O (0.55) and dashed lines indicating super-solar C/O (1). Lower-metallicity models fall below the plotted abundance range. The pressure axis is restricted to the approximate region probed, estimated from pressure contribution functions. The full-spectrum atm+SC retrieval yields CO$_2$ and CO abundances above the low-metallicity equilibrium expectations, suggesting disequilibrium chemistry would be required. In contrast, the truncated atm+SC retrieval yields lower abundances more consistent with equilibrium predictions, highlighting that conclusions about chemical disequilibrium are sensitive to the wavelength range included in the retrieval.
    }
    \label{fig:co2_co_vulcan}
\end{figure*}

The atm+SC solution also points to possible chemical disequilibrium. At the equilibrium temperature of TOI-3235~b (604~K), carbon is expected to reside predominantly in CH$_4$, and both CO and CO$_2$ are expected to decrease steeply with decreasing atmospheric metallicity (see Section~\ref{app:A}). Our VULCAN equilibrium calculations therefore predict very low CO and CO$_2$ abundances for sub-solar metallicities. In contrast, the full-spectrum atm+SC posterior places substantial probability at CO and CO$_2$ abundances orders of magnitude higher than these equilibrium expectations. Figure~\ref{fig:co2_co_vulcan} compares the atm+SC posterior distributions for CO$_2$ and CO with VULCAN chemical equilibrium predictions at both solar C/O (0.55) and super-solar C/O (1). For clarity, we show only the 1, 10, and 100$\times$ solar metallicity models; lower-metallicity models fall below the plotted abundance range. Under the full-spectrum atm+SC interpretation, the elevated CO$_2$ and CO abundances would require disequilibrium chemistry to reproduce retrieved abundances. However, the truncated atm+SC retrieval yields lower CO$_2$ and CO abundances that may be consistent with equilibrium expectations, as shown by the orange posteriors in Figure~\ref{fig:co2_co_vulcan}. The differing results between the full-spectrum and truncated retrievals highlight the uncertainties induced by unmodeled correlated noise, particularly at wavelengths beyond 4.7~$\mu$m, and prevent us from drawing robust conclusions about chemical disequilibrium.

As further evidence for caution, the atm+SC+GP retrieval shows that wavelength-correlated residual structure remains after fitting the deterministic atmosphere+SC model, suggesting physical model inadequacy. Furthermore, the truncated atm+SC retrieval, which excludes data beyond 4.7~$\mu$m, shifts the metallicity and C/O posteriors to different values, indicating sensitivity to the reddest wavelengths where residual structure is most prominent. We therefore interpret the low metallicity and possible disequilibrium chemistry as conditional implications of the atm+SC model, rather than as robust atmospheric measurements.

\subsection{Emission spectroscopy as a path forward}
\label{sec:discussion_emission}

Emission spectroscopy provides a pathway to break the model-dependent degeneracies present in our transmission analysis. Unlike transmission spectroscopy, secondary eclipse observations are much less sensitive to unocculted stellar heterogeneities, enabling a more direct probe of the planetary atmosphere without the confounding effects of stellar contamination.

To determine the conditions in which emission spectroscopy can constrain the atmospheric metallicity of TOI-3235~b, we generate a grid of synthetic emission spectra spanning a range of injected metallicities and cloud-top pressures. We take the atm+SC+GP results as described in Section~\ref{subsec:GP_results}, divide the metallicity distribution into equally spaced bins, and extract the median molecular abundances from each bin. We then systematically vary the cloud-top pressure across these metallicity values and generate forward model emission spectra using \texttt{POSEIDON}. The cloud-top pressure grid spans $\log P_{\rm cloud} = -6$ to $3$~bar in six equally spaced steps, covering the full range of our prior. To restrict the experiment to a two-dimensional grid in metallicity and cloud-top pressure, we fix the haze parameters (amplitude and scattering slope) to their median retrieved values from the atm+SC+GP posterior. The forward models are computed at a resolution of $R = 20{,}000$ over a wavelength range of $0.5$--$5.7~\mu$m, then binned to $R = 100$ to match the transmission spectrum. We adopt a two-parameter gradient pressure-temperature profile, with the upper atmosphere temperature set to 400~K and the deep atmosphere temperature set to 800~K, bracketing the planet's equilibrium temperature of 604~K. This parameterization is motivated by the thermal structure models of \citet{molliere2015model}. To estimate realistic uncertainties, we propagate the photometric precision from our observed transmission light curve residuals using
\begin{equation}
    \sigma_{\rm eclipse} = \sigma_{\rm lc} \sqrt{\frac{1}{N_{\rm out}} + \frac{1}{N_{\rm in}}},
\end{equation}
where $\sigma_{\rm lc}$ is the standard deviation of the transmission light curve residuals at each wavelength, and $N_{\rm in} \approx N_{\rm out} \approx 7200$ integrations, assuming the eclipse duration equals the transit duration of 1.48 hours. This corresponds to a single NIRSpec/PRISM eclipse observation with the same instrumental configuration as our GO~3731 transit observation, yielding a median precision of $\sim$98~ppm per spectral bin.

We retrieve each synthetic emission spectrum using \texttt{POSEIDON}, closely following the framework as the atmosphere-only transmission retrieval described in Section~\ref{subsec:atmosphere_setup}. The emission retrievals use identical molecular species, abundance priors, cloud parameterization, and sampling configuration. The key difference is the use of a two-parameter gradient pressure-temperature profile, with uniform priors between 400--1000~K for both the upper and deep atmosphere temperatures, constrained such that the upper atmosphere temperature remains lower than the deep atmosphere temperature.

For each retrieval, we evaluate two metrics to assess the constraining power of emission spectroscopy. First, we compute $W_{68}$, the width of the central 68\% credible interval of the prior-corrected retrieved metallicity posterior, as a measure of metallicity precision. Second, we compute $P_{\rm correct}$, the posterior probability assigned to the correct classification of the injected atmosphere as either sub-solar or super-solar in metallicity. For sub-solar injections, $P_{\rm correct}$ is the integral of the posterior from $-\infty$ to 0; for super-solar injections, it is the integral from 0 to $+\infty$.

\begin{figure*}
    \centering
    \includegraphics[width=0.8\linewidth]{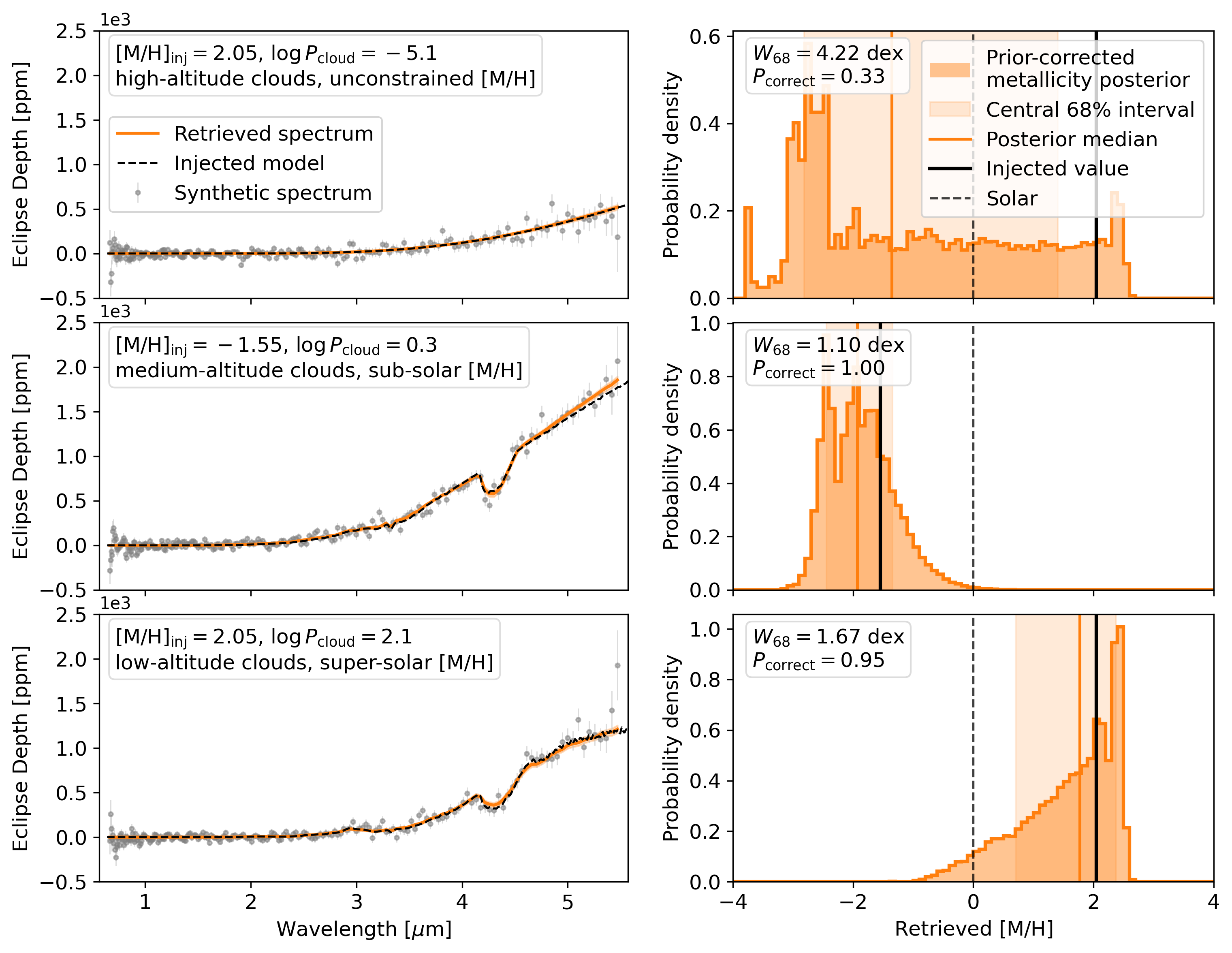}
    \caption{Three representative cases from our synthetic emission retrieval experiment. The left panels show the synthetic emission spectrum (gray points), the injected forward model (black dashed line), and the retrieved spectrum (orange line and shaded $1\sigma$ region). The right panels show the prior-corrected metallicity posterior (orange histogram), with the injected value (black solid line), posterior median (orange solid line), central 68\% credible interval (orange shaded region), and solar metallicity (black dashed line) indicated. The metrics $W_{68}$ and $P_{\rm correct}$ are annotated in each panel. High-altitude clouds (top row) yield broad, uninformative metallicity posteriors, while deeper cloud decks (middle and bottom rows) yield well-constrained metallicity estimates.}
    \label{fig:emission_example_spectra}
\end{figure*}

Figure~\ref{fig:emission_example_spectra} shows three cases from our synthetic emission retrieval experiment. The top row shows a case with high-altitude clouds ($\log P_{\rm cloud} = -5.1$~bar), where the emission spectrum is largely featureless and the retrieved metallicity posterior is broad ($W_{68} = 4.2$~dex) with low classification confidence ($P_{\rm correct} = 0.33$). The middle row shows a case with moderate cloud-top pressure ($\log P_{\rm cloud} = 0.3$~bar), where molecular features emerge and the metallicity is well constrained ($W_{68} = 1.10$~dex, $P_{\rm correct} = 1.00$). The bottom row shows a case with a deep cloud deck ($\log P_{\rm cloud} = 2.1$~bar) and super-solar metallicity, which is confidently recovered ($W_{68} = 1.67$~dex, $P_{\rm correct} = 0.95$). These cloud-top pressure regimes are defined relative to the photospheric pressures probed by emission spectroscopy; clouds above this level obscure molecular features, while clouds below have minimal impact on the spectrum. These examples illustrate how the presence or absence of high-altitude clouds impacts whether emission spectroscopy can successfully constrain the atmospheric metallicity.

\begin{figure*}
    \centering
    \includegraphics[width=\linewidth]{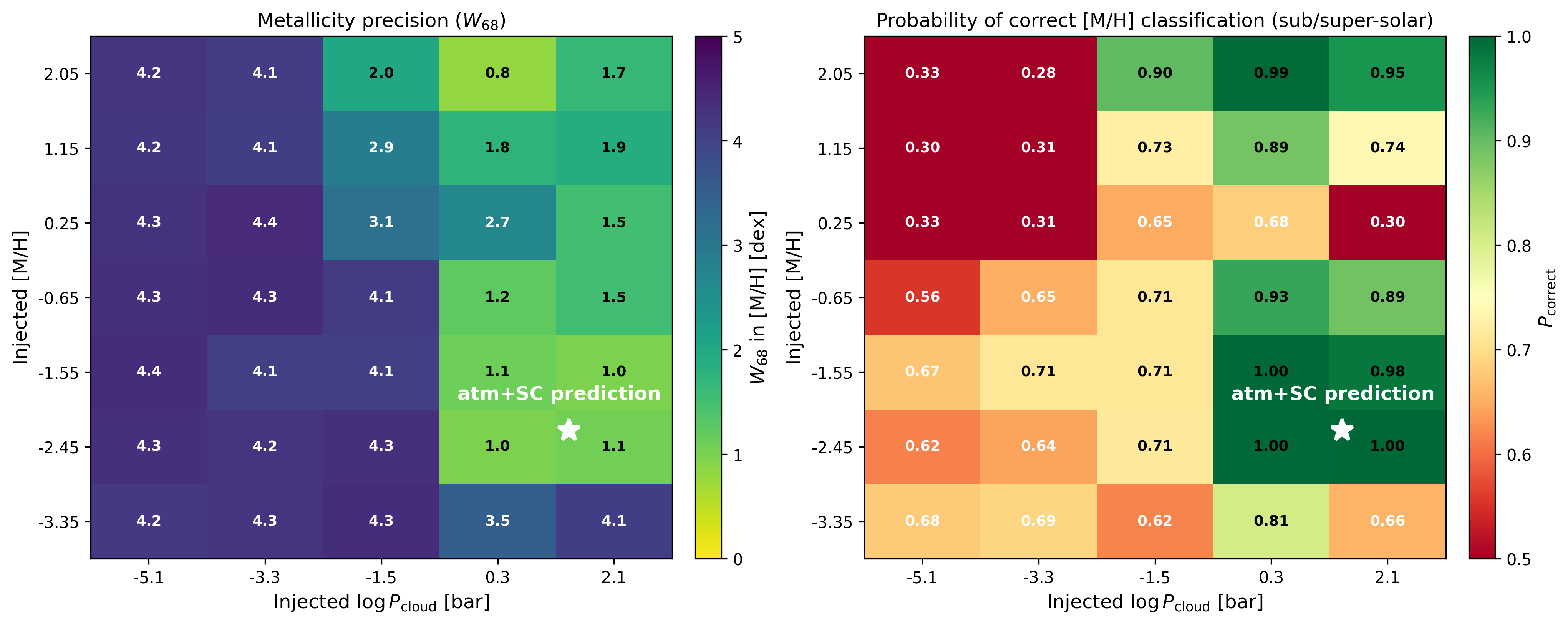}
    \caption{Emission retrieval performance across a grid of injected atmospheric metallicities and cloud-top pressures. The left panel shows $W_{68}$, the width of the central 68\% credible interval of the retrieved metallicity posterior, indicating metallicity precision. The right panel shows $P_{\rm correct}$, the posterior probability of correctly classifying the atmosphere as sub-solar or super-solar. Deeper cloud decks (higher $\log P_{\rm cloud}$) allow for stronger molecular features, leading to narrower metallicity posteriors and higher classification accuracy. The white star marks the location corresponding to the atm+SC retrieval prediction, which falls in a regime where emission spectroscopy would provide a confident metallicity constraint.}
    \label{fig:emission_metallicity_grid}
\end{figure*}

Figure~\ref{fig:emission_metallicity_grid} summarizes the results across our full grid of injected metallicities and cloud-top pressures. The left panel shows the metallicity precision $W_{68}$, and the right panel shows the classification probability $P_{\rm correct}$. Both metrics improve as the cloud deck moves to higher pressures, allowing for stronger molecular features that yield tighter metallicity constraints. Notably, if the atm+SC retrieval solution were representative of the true atmosphere (sub-solar metallicity with a moderate cloud deck), this scenario falls within the regime where emission spectroscopy would confidently identify the metallicity as sub-solar with a single observation.

These simulations demonstrate that an emission spectrum of TOI-3235~b would yield one of two informative outcomes. In the first scenario, if the cloud deck lies at moderate to high pressures, molecular features would be detectable and the atmospheric metallicity would be well constrained. This would test whether the sub-solar metallicity suggested by the atm+SC transmission retrieval is genuine and enable physical implications for the planet's formation history. In the second scenario, if high-altitude clouds obscure atmospheric features, the emission spectrum would appear featureless. This outcome would indicate that the molecular features observed in the transmission spectrum are not of planetary origin, confirming that the retrievals without a GP are biased by stellar contamination. This would also yield a rich empirical stellar contamination spectrum of an M dwarf system, providing valuable constraints for future studies of stellar heterogeneity effects in transmission spectroscopy. Thus, with a single emission observation, we would either constrain the atmospheric metallicity of TOI-3235~b and its implications for giant planet formation around low-mass stars, or advance our understanding of stellar contamination in M dwarf systems.

\section{Conclusion} \label{sec:conclusion}

In this work, we present a detailed atmospheric characterization study of TOI-3235~b, a $0.665~M_J$ giant planet orbiting a $0.39~M_\odot$ M dwarf, to understand how its atmospheric composition may connect to possible formation scenarios. We apply a hierarchical retrieval framework incorporating an atmospheric model, a parametric stellar contamination model, and a Gaussian process to disentangle planetary and stellar signals. The inferred atmospheric properties are model-dependent, with different retrieval configurations yielding different constraints on molecular abundances and metallicity. The atm+SC+GP retrieval yields the broadest posterior distributions, reflecting uncertainties from residual wavelength-correlated structure that would otherwise be absorbed into the physical model parameters. This makes it the most conservative interpretation of our data, avoiding over-precise and potentially biased conclusions about the atmospheric properties of TOI-3235~b.

If the deterministic atm+SC retrieval is representative of the true atmosphere, TOI-3235~b would have a CH$_4$ posterior with significant probability away from the prior bounds and a sub-solar atmospheric metallicity. Such a low metallicity would challenge standard expectations from both core accretion and gravitational instability for giant planet formation around M dwarfs. Additionally, elevated CO and CO$_2$ abundances relative to chemical equilibrium predictions could require disequilibrium processes. However, these results should be interpreted with caution, as the atm+SC model leaves wavelength-correlated residual structure that suggests physical model inadequacy.

A single eclipse observation could clarify these model-dependent degeneracies by probing the planetary atmosphere largely free of stellar contamination. If the cloud deck lies at moderate to high pressures, emission spectroscopy would constrain the atmospheric metallicity and enable physical implications for giant planet formation around low-mass stars. If high-altitude clouds obscure atmospheric features, this would confirm that stellar contamination dominates the transmission spectrum and yield a rich empirical stellar contamination spectrum of an M dwarf system. Either outcome would advance our understanding of giant planet formation around M dwarfs or stellar contamination characterization in M dwarf systems, knowledge that is essential for accurately characterizing the large population of planets orbiting these common hosts.

\section{Data and Figures Availability} \label{sec:data_availability}
The data presented in this paper are from JWST GO 3731 and are available at the Mikulski Archive for Space Telescopes (MAST) at the Space Telescope Science Institute. The specific observation can be accessed via \dataset[doi:10.17909/02a7-ck05]{https://doi.org/10.17909/02a7-ck05}.

Main data products and models are available on Zenodo at \dataset[doi:10.5281/zenodo.20738525]{https://doi.org/10.5281/zenodo.20738525}. The code to recreate corresponding figures is available on GitHub at \url{https://github.com/zoekko/TOI3235b-JWST-transmission}.

\begin{acknowledgments}

We thank the anonymous referee for their insightful comments, which have greatly improved the manuscript. This work is based in part on observations made with the NASA/ESA/CSA James Webb Space Telescope. The data were obtained from the Mikulski Archive for Space Telescopes at the Space Telescope Science Institute, which is operated by the Association of Universities for Research in Astronomy, Inc., under NASA contract NAS 5-03127 for JWST. These observations are associated with program \#3731. Support for program \#3731 was provided by NASA through a grant from the Space Telescope Science Institute, which is operated by the Association of Universities for Research in Astronomy, Inc., under NASA contract NAS 5-03127. A.J.\ acknowledges support from Fondecyt project 1251439. 

This work has been carried out within the framework of the NCCR PlanetS supported by the Swiss National Science Foundation under grant 51NF40\_205606.
R.Bu.\ acknowledges the financial support from the Poincar\'e fellowship of the Observatoire de la C\^ote d'Azur and from the ERC Advanced Grant ``TiPPi -- Turbulence, Pebbles and Planetesimals: The Origin of Minor Bodies in the Solar System'' (project ID 855130, PI: Klahr).
R.Br.\ acknowledges support from FONDECYT Project 1241963.

\end{acknowledgments}

\facility{James Webb Space Telescope}

\software{
\texttt{george} \citep{george},
\texttt{celerite} \citep{celerite},
\texttt{transitspectroscopy} \citep{transitspectroscopy},
\texttt{juliet} \citep{juliet},
\texttt{POSEIDON} \citep{macdonald2023poseidon},
\texttt{dynesty} \citep{speagle2020dynesty},
\texttt{VULCAN} \citep{vulcan},
\texttt{pysynphot} \citep{lim2015pysynphot}
}
\bibliography{references}{}

@ARTICLE{bryant2025,
       author = {{Bryant}, Edward M. and {Jord{\'a}n}, Andr{\'e}s and {Hartman}, Joel D. and {Bayliss}, Daniel and {Sedaghati}, Elyar and {Barkaoui}, Khalid and {Chouqar}, Jamila and {Pozuelos}, Francisco J. and {Thorngren}, Daniel P. and {Timmermans}, Mathilde and {Almenara}, Jose Manuel and {Chilingarian}, Igor V. and {Collins}, Karen A. and {Gan}, Tianjun and {Howell}, Steve B. and {Narita}, Norio and {Palle}, Enric and {Rackham}, Benjamin V. and {Triaud}, Amaury H.~M.~J. and {Bakos}, Gaspar {\'A}. and {Brahm}, Rafael and {Hobson}, Melissa J. and {Van Eylen}, Vincent and {Amado}, Pedro J. and {Arnold}, Luc and {Bonfils}, Xavier and {Burdanov}, Artem and {Cadieux}, Charles and {Caldwell}, Douglas A. and {Casanova}, Victor and {Charbonneau}, David and {Clark}, Catherine A. and {Collins}, Kevin I. and {Daylan}, Tansu and {Dransfield}, Georgina and {Demory}, Brice-Olivier and {Ducrot}, Elsa and {Fern{\'a}ndez-Rodr{\'\i}guez}, Gareb and {Fukuda}, Izuru and {Fukui}, Akihiko and {Gillon}, Micha{\"e}l and {Gore}, Rebecca and {Hooton}, Matthew J. and {Ikuta}, Kai and {Jehin}, Emmanuel and {Jenkins}, Jon M. and {Levine}, Alan M. and {Littlefield}, Colin and {Murgas}, Felipe and {Nguyen}, Kendra and {Parviainen}, Hannu and {Queloz}, Didier and {Seager}, S. and {Sebastian}, Daniel and {Srdoc}, Gregor and {Vanderspek}, R. and {Winn}, Joshua N. and {de Wit}, Julien and {Z{\'u}{\~n}iga-Fern{\'a}ndez}, Sebasti{\'a}n},
        title = "{A transiting giant planet in orbit around a 0.2-solar-mass host star}",
      journal = {Nature Astronomy},
         year = 2025,
        month = jul,
       volume = {9},
        pages = {1031-1044},
          doi = {10.1038/s41550-025-02552-4},
archivePrefix = {arXiv},
       eprint = {2506.07931},
 primaryClass = {astro-ph.EP},
       adsurl = {https://ui.adsabs.harvard.edu/abs/2025NatAs...9.1031B}
}

@ARTICLE{McCreery:2025,
       author = {{McCreery}, Patrick and {Dos Santos}, Leonardo A. and {Espinoza}, N{\'e}stor and {Allart}, Romain and {Kirk}, James},
        title = "{Tracing the Winds: A Uniform Interpretation of Helium Escape in Exoplanets from Archival Spectroscopic Observations}",
      journal = {\apj},
         year = 2025,
        month = feb,
       volume = {980},
       number = {1},
          eid = {125},
        pages = {125},
          doi = {10.3847/1538-4357/ada6b9},
archivePrefix = {arXiv},
       eprint = {2501.03998},
 primaryClass = {astro-ph.EP},
       adsurl = {https://ui.adsabs.harvard.edu/abs/2025ApJ...980..125M}
}

@ARTICLE{EP:2025,
       author = {{Espinoza}, N{\'e}stor and {Perrin}, Marshall D.},
        title = "{Highlights from Exoplanet Observations by the James Webb Space Telescope}",
      journal = {arXiv e-prints},
         year = 2025,
        month = may,
          eid = {arXiv:2505.20520},
        pages = {arXiv:2505.20520},
          doi = {10.48550/arXiv.2505.20520},
archivePrefix = {arXiv},
       eprint = {2505.20520},
 primaryClass = {astro-ph.EP},
       adsurl = {https://ui.adsabs.harvard.edu/abs/2025arXiv250520520E}
}

@ARTICLE{espinoza:2017,
       author = {{Espinoza}, N{\'e}stor and {Fortney}, Jonathan J. and {Miguel}, Yamila and {Thorngren}, Daniel and {Murray-Clay}, Ruth},
        title = "{Metal Enrichment Leads to Low Atmospheric C/O Ratios in Transiting Giant Exoplanets}",
      journal = {\apjl},
         year = 2017,
        month = mar,
       volume = {838},
       number = {1},
          eid = {L9},
        pages = {L9},
          doi = {10.3847/2041-8213/aa65ca},
archivePrefix = {arXiv},
       eprint = {1611.08616},
 primaryClass = {astro-ph.EP},
       adsurl = {https://ui.adsabs.harvard.edu/abs/2017ApJ...838L...9E}
}

@ARTICLE{mordasini:2016,
       author = {{Mordasini}, C. and {van Boekel}, R. and {Molli{\`e}re}, P. and {Henning}, Th. and {Benneke}, Bj{\"o}rn},
        title = "{The Imprint of Exoplanet Formation History on Observable Present-day Spectra of Hot Jupiters}",
      journal = {\apj},
         year = 2016,
        month = nov,
       volume = {832},
       number = {1},
          eid = {41},
        pages = {41},
          doi = {10.3847/0004-637X/832/1/41},
archivePrefix = {arXiv},
       eprint = {1609.03019},
 primaryClass = {astro-ph.EP},
       adsurl = {https://ui.adsabs.harvard.edu/abs/2016ApJ...832...41M}
}

@ARTICLE{almenara2024,
       author = {{Almenara}, J.~M. and {Bonfils}, X. and {Bryant}, E.~M. and {Jord{\'a}n}, A. and {H{\'e}brard}, G. and {Martioli}, E. and {Correia}, A.~C.~M. and {Astudillo-Defru}, N. and {Cadieux}, C. and {Arnold}, L. and {Artigau}, {\'E}. and {Bakos}, G. {\'A}. and {Barros}, S.~C.~C. and {Bayliss}, D. and {Bouchy}, F. and {Bou{\'e}}, G. and {Brahm}, R. and {Carmona}, A. and {Charbonneau}, D. and {Ciardi}, D.~R. and {Cloutier}, R. and {Cointepas}, M. and {Cook}, N.~J. and {Cowan}, N.~B. and {Delfosse}, X. and {Dias do Nascimento}, J. and {Donati}, J.-F. and {Doyon}, R. and {Forveille}, T. and {Fouqu{\'e}}, P. and {Gaidos}, E. and {Gilbert}, E.~A. and {Gomes da Silva}, J. and {Hartman}, J.~D. and {Hesse}, K. and {Hobson}, M.~J. and {Jenkins}, J.~M. and {Kiefer}, F. and {Kostov}, V.~B. and {Laskar}, J. and {Lendl}, M. and {L'Heureux}, A. and {Martins}, J.~H.~C. and {Menou}, K. and {Moutou}, C. and {Murgas}, F. and {Polanski}, A.~S. and {Rapetti}, D. and {Sedaghati}, E. and {Shang}, H.},
        title = "{TOI-4860 b, a short-period giant planet transiting an M3.5 dwarf}",
      journal = {\aap},
         year = 2024,
        month = mar,
       volume = {683},
          eid = {A166},
        pages = {A166},
          doi = {10.1051/0004-6361/202346999},
archivePrefix = {arXiv},
       eprint = {2308.01454},
 primaryClass = {astro-ph.EP},
       adsurl = {https://ui.adsabs.harvard.edu/abs/2024A&A...683A.166A}
}

@ARTICLE{almenara2022,
       author = {{Almenara}, J.~M. and {Bonfils}, X. and {Forveille}, T. and {Astudillo-Defru}, N. and {Ciardi}, D.~R. and {Schwarz}, R.~P. and {Collins}, K.~A. and {Cointepas}, M. and {Lund}, M.~B. and {Bouchy}, F. and {Charbonneau}, D. and {D{\'\i}az}, R.~F. and {Delfosse}, X. and {Kidwell}, R.~C. and {Kunimoto}, M. and {Latham}, D.~W. and {Lissauer}, J.~J. and {Murgas}, F. and {Ricker}, G. and {Seager}, S. and {Vezie}, M. and {Watanabe}, D.},
        title = "{TOI-3884 b: A rare 6-R$_{E}$ planet that transits a low-mass star with a giant and likely polar spot}",
      journal = {\aap},
         year = 2022,
        month = nov,
       volume = {667},
          eid = {L11},
        pages = {L11},
          doi = {10.1051/0004-6361/202244791},
archivePrefix = {arXiv},
       eprint = {2210.10909},
 primaryClass = {astro-ph.EP},
       adsurl = {https://ui.adsabs.harvard.edu/abs/2022A&A...667L..11A}
}

@misc{transitspectroscopy,
  author       = {Espinoza, Nestor},
  title        = {TransitSpectroscopy},
  month        = aug,
  year         = 2022,
  publisher    = {Zenodo},
  version      = {0.3.11},
  doi          = {10.5281/zenodo.6960924},
  url          = {https://doi.org/10.5281/zenodo.6960924}
}

@article{juliet,
       author = {{Espinoza}, N{\'e}stor and {Kossakowski}, Diana and {Brahm}, Rafael},
        title = "{juliet: a versatile modelling tool for transiting and non-transiting exoplanetary systems}",
      journal = {\mnras},
         year = "2019",
        month = "Dec",
       volume = {490},
       number = {2},
        pages = {2262-2283},
          doi = {10.1093/mnras/stz2688},
archivePrefix = {arXiv},
       eprint = {1812.08549},
 primaryClass = {astro-ph.EP},
       adsurl = {https://ui.adsabs.harvard.edu/abs/2019MNRAS.490.2262E}
}

@article{rackham2018,
  title={The transit light source effect: false spectral features and incorrect densities for M-dwarf transiting planets},
  author={Rackham, Benjamin V and Apai, D{\'a}niel and Giampapa, Mark S},
  journal={The Astrophysical Journal},
  volume={853},
  number={2},
  pages={122},
  year={2018},
  publisher={IOP Publishing}
}

@article{smitha2024first,
  title={First Calculations of Starspot Spectra Based on 3D Radiative Magnetohydrodynamics Simulations},
  author={Smitha, HN and Shapiro, Alexander I and Witzke, Veronika and Kostogryz, Nadiia M and Unruh, Yvonne C and Bhatia, Tanayveer S and Cameron, Robert and Seager, Sara and Solanki, Sami K},
  journal={The Astrophysical Journal Letters},
  volume={978},
  number={1},
  pages={L13},
  year={2024},
  publisher={IOP Publishing}
}

@article{des2008rayleigh,
  title={Rayleigh scattering by H2 in the extrasolar planet HD 209458b},
  author={Des Etangs, A Lecavelier and Vidal-Madjar, A and D{\'e}sert, J-M and Sing, D},
  journal={Astronomy \& Astrophysics},
  volume={485},
  number={3},
  pages={865--869},
  year={2008},
  publisher={EDP Sciences}
}

@article{kipping2025exoplaneteers,
  title={Exoplaneteers Keep Overestimating Sigma Significances},
  author={Kipping, David and Benneke, Bj{\"o}rn},
  journal={arXiv preprint arXiv:2506.05392},
  year={2025}
}

@article{mccullough2014water,
  title={Water vapor in the spectrum of the extrasolar planet HD 189733b. I. The transit},
  author={McCullough, PR and Crouzet, N and Deming, D and Madhusudhan, N},
  journal={The Astrophysical Journal},
  volume={791},
  number={1},
  pages={55},
  year={2014},
  publisher={IOP Publishing}
}

@article{welbanks2019degeneracies,
  title={On degeneracies in retrievals of exoplanetary transmission spectra},
  author={Welbanks, Luis and Madhusudhan, Nikku},
  journal={The Astronomical Journal},
  volume={157},
  number={5},
  pages={206},
  year={2019},
  publisher={American Astronomical Society}
}

@article{macdonald2023poseidon,
  title={POSEIDON: a multidimensional atmospheric retrieval code for exoplanet spectra},
  author={MacDonald, Ryan J},
  journal={arXiv preprint arXiv:2410.18181},
  year={2023}
}

@article{seager2000theoretical,
  title={Theoretical Transmission Spectra during Extrasolar Giant PlanetTransits},
  author={Seager, SV and Sasselov, Dimitar D},
  journal={The Astrophysical Journal},
  volume={537},
  number={2},
  pages={916},
  year={2000},
  publisher={IOP Publishing}
}

@article{schlecker2022,
  title={RV-detected planets around M dwarfs: Challenges for core accretion models},
  author={Schlecker, Martin and Burn, Remo and Sabotta, Silvia and Seifert, Antonia and Henning, Th and Emsenhuber, Alexandre and Mordasini, Christoph and Reffert, Sabine and Shan, Yutong and Klahr, Hubert},
  journal={Astronomy \& Astrophysics},
  volume={664},
  pages={A180},
  year={2022},
  publisher={EDP Sciences}
}

@article{rathcke2021hst,
  title={HST PanCET Program: a complete near-UV to infrared transmission spectrum for the hot Jupiter WASP-79b},
  author={Rathcke, Alexander D and MacDonald, Ryan J and Barstow, Joanna K and Goyal, Jayesh M and Lopez-Morales, Mercedes and Mendon{\c{c}}a, Jo{\~a}o M and Sanz-Forcada, Jorge and Henry, Gregory W and Sing, David K and Alam, Munazza K and others},
  journal={The Astronomical Journal},
  volume={162},
  number={4},
  pages={138},
  year={2021},
  publisher={IOP Publishing}
}

@article{bennett2025additional,
  title={Additional JWST/NIRSpec Transits of the Rocky M Dwarf Exoplanet GJ 1132 b Reveal a Featureless Spectrum},
  author={Bennett, Katherine A and MacDonald, Ryan J and Peacock, Sarah and Perez, Junellie and May, EM and Moran, Sarah E and Alderson, Lili and Lustig-Yaeger, Jacob and Wakeford, Hannah R and Sing, David K and others},
  journal={arXiv preprint arXiv:2508.10579},
  year={2025}
}

@article{fournier2024near,
  title={Near-infrared transmission spectroscopy of HAT-P-18 b with NIRISS: Disentangling planetary and stellar features in the era of JWST},
  author={Fournier-Tondreau, Marylou and MacDonald, Ryan J and Radica, Michael and Lafreni{\`e}re, David and Welbanks, Luis and Piaulet, Caroline and Coulombe, Louis-Philippe and Allart, Romain and Morel, Kim and Artigau, {\'E}tienne and others},
  journal={Monthly Notices of the Royal Astronomical Society},
  volume={528},
  number={2},
  pages={3354--3377},
  year={2024},
  publisher={Oxford University Press}
}

@article{bushouse2023jwst,
  title={JWST calibration pipeline},
  author={Bushouse, Howard and Eisenhamer, Jonathan and Dencheva, Nadia and Davies, James and Greenfield, Perry and Morrison, Jane and Hodge, Phil and Simon, Bernie and Grumm, David and Droettboom, Michael and others},
  journal={Zenodo},
  year={2023}
}

@ARTICLE{george,
        author = {{Ambikasaran}, Sivaram and {Foreman-Mackey}, Daniel and {Greengard}, Leslie and {Hogg}, David W. and {O'Neil}, Michael},
         title = "{Fast Direct Methods for Gaussian Processes}",
       journal = {IEEE Transactions on Pattern Analysis and Machine Intelligence},
          year = 2015,
         month = jun,
        volume = {38},
         pages = {252},
           doi = {10.1109/TPAMI.2015.2448083},
 archivePrefix = {arXiv},
        eprint = {1403.6015},
  primaryClass = {math.NA},
        adsurl = {https://ui.adsabs.harvard.edu/abs/2015ITPAM..38..252A}
}

@article{kipping2013efficient,
  title={Efficient, uninformative sampling of limb darkening coefficients for two-parameter laws},
  author={Kipping, David M},
  journal={Monthly Notices of the Royal Astronomical Society},
  volume={435},
  number={3},
  pages={2152--2160},
  year={2013},
  publisher={Oxford University Press}
}

@article{coulombe2024biases,
  title={Biases in Exoplanet Transmission Spectra Introduced by Limb-darkening Parametrization},
  author={Coulombe, Louis-Philippe and Roy, Pierre-Alexis and Benneke, Bj{\"o}rn},
  journal={The Astronomical Journal},
  volume={168},
  number={5},
  pages={227},
  year={2024},
  publisher={IOP Publishing}
}

@article{celerite,
    author = {{Foreman-Mackey}, D. and {Agol}, E. and {Angus}, R. and
              {Ambikasaran}, S.},
     title = {Fast and scalable Gaussian process modeling
              with applications to astronomical time series},
      year = {2017},
   journal = {AJ},
    volume = {154},
     pages = {220},
       doi = {10.3847/1538-3881/aa9332},
       url = {https://arxiv.org/abs/1703.09710}
}

@article{guilluy2024gaps,
  title={The GAPS Programme at TNG-LIV. A He I survey of close-in giant planets hosted by MK dwarf stars with GIANO-B},
  author={Guilluy, G and D’Arpa, MC and Bonomo, ALDO STEFANO and Spinelli, R and Biassoni, F and Fossati, L and Maggio, Antonio and Giacobbe, Paolo and Lanza, Antonino Francesco and Sozzetti, Alessandro and others},
  journal={Astronomy \& Astrophysics},
  volume={686},
  pages={A83},
  year={2024},
  publisher={EDP Sciences}
}

@article{witzke2022can,
  title={Can 1D radiative-equilibrium models of faculae be used for calculating contamination of transmission spectra?},
  author={Witzke, Veronika and Shapiro, Alexander I and Kostogryz, Nadiia M and Cameron, Robert and Rackham, Benjamin V and Seager, Sara and Solanki, Sami K and Unruh, Yvonne C},
  journal={The Astrophysical Journal Letters},
  volume={941},
  number={2},
  pages={L35},
  year={2022},
  publisher={IOP Publishing}
}

@article{norris2023spectral,
  title={Spectral variability of photospheric radiation due to faculae--II. Facular contrasts for cool main-sequence stars},
  author={Norris, Charlotte M and Unruh, Yvonne C and Witzke, Veronika and Solanki, Sami K and Krivova, Natalie A and Shapiro, Alexander I and Yeo, Kok Leng and Cameron, Robert and Beeck, Benjamin},
  journal={Monthly Notices of the Royal Astronomical Society},
  volume={524},
  number={1},
  pages={1139--1155},
  year={2023},
  publisher={Oxford University Press}
}

@misc{mercier2025,
      title={What's in Your Transit? Towards Reliably Getting $5\times$ More Science from Exoplanet Transit Data}, 
      author={Samson J. Mercier and Julien de Wit and Benjamin V. Rackham},
      year={2025},
      eprint={2510.00124},
      archivePrefix={arXiv},
      primaryClass={astro-ph.EP},
      url={https://arxiv.org/abs/2510.00124}, 
}

@article{espinoza2025dreams,
   title={JWST-TST DREAMS: NIRSpec/PRISM Transmission Spectroscopy of the Habitable Zone Planet TRAPPIST-1 e},
   volume={990},
   ISSN={2041-8213},
   number={2},
   journal={The Astrophysical Journal Letters},
   publisher={American Astronomical Society},
   author={Espinoza, Néstor and Allen, Natalie H. and Glidden, Ana and Lewis, Nikole K. and Seager, Sara and Cañas, Caleb I. and Grant, David and Gressier, Amélie and Courreges, Shelby and Stevenson, Kevin B. and Ranjan, Sukrit and Colón, Knicole and Morris, Brett M. and MacDonald, Ryan J. and Long, Douglas and Wakeford, Hannah R. and Valenti, Jeff A. and Alderson, Lili and Batalha, Natasha E. and Challener, Ryan C. and Huang, Jingcheng and Lin, Zifan and Louie, Dana R. and Mullens, Elijah and Valentine, Daniel and Mountain, C. Matt and Pueyo, Laurent and Perrin, Marshall D. and Bellini, Andrea and Kammerer, Jens and Libralato, Mattia and Rebollido, Isabel and Rickman, Emily and Sohn, Sangmo Tony and van der Marel, Roeland P.},
   year={2025},
   month=sep, pages={L52} }

@article{laughlin2004,
  title={The core accretion model predicts few Jovian-mass planets orbiting red dwarfs},
  author={Laughlin, Gregory and Bodenheimer, Peter and Adams, Fred C},
  journal={The Astrophysical Journal},
  volume={612},
  number={1},
  pages={L73},
  year={2004},
  publisher={IOP Publishing}
}

@article{burn2021,
  title={The new generation planetary population synthesis (ngpps)-iv. planetary systems around low-mass stars},
  author={Burn, Remo and Schlecker, Martin and Mordasini, Christoph and Emsenhuber, Alexandre and Alibert, Yann and Henning, Thomas and Klahr, Hubert and Benz, Willy},
  journal={Astronomy \& Astrophysics},
  volume={656},
  pages={A72},
  year={2021},
  publisher={EDP Sciences}
}

@article{kanodia2023,
  title={TOI-5205b: a short-period Jovian planet transiting a mid-M dwarf},
  author={Kanodia, Shubham and Mahadevan, Suvrath and Libby-Roberts, Jessica and Stefansson, Gudmundur and Ca{\~n}as, Caleb I and Piette, Anjali AA and Boss, Alan and Teske, Johanna and Chambers, John and Zeimann, Greg and others},
  journal={The Astronomical Journal},
  volume={165},
  number={3},
  pages={120},
  year={2023},
  publisher={IOP Publishing}
}

@article{liu2019,
  title={Super-Earth masses sculpted by pebble isolation around stars of different masses},
  author={Liu, Beibei and Lambrechts, Michiel and Johansen, Anders and Liu, Fan},
  journal={Astronomy \& Astrophysics},
  volume={632},
  pages={A7},
  year={2019},
  publisher={EDP Sciences}
}

@article{hobson2023,
  title={TOI-3235 b: a transiting giant planet around an M4 dwarf star},
  author={Hobson, Melissa J and Jord{\'a}n, Andr{\'e}s and Bryant, EM and Brahm, R and Bayliss, D and Hartman, JD and Bakos, G{\'A} and Henning, Th and Almenara, Jose Manuel and Barkaoui, Khalid and others},
  journal={The Astrophysical Journal Letters},
  volume={946},
  number={1},
  pages={L4},
  year={2023},
  publisher={IOP Publishing}
}

@article{molliere2015model,
  title={Model atmospheres of irradiated exoplanets: the influence of stellar parameters, metallicity, and the C/O ratio},
  author={Molli{\`e}re, Paul and van Boekel, Roy and Dullemond, C and Henning, Th and Mordasini, Christoph},
  journal={The Astrophysical Journal},
  volume={813},
  number={1},
  pages={47},
  year={2015},
  publisher={IOP Publishing}
}

@article{oberg2011,
  title={The effects of snowlines on C/O in planetary atmospheres},
  author={{\"O}berg, Karin I and Murray-Clay, Ruth and Bergin, Edwin A},
  journal={The Astrophysical Journal Letters},
  volume={743},
  number={1},
  pages={L16},
  year={2011},
  publisher={IOP Publishing}
}

@article{madhusudhan2019,
  title={Exoplanetary atmospheres: key insights, challenges, and prospects},
  author={Madhusudhan, Nikku},
  journal={Annual Review of Astronomy and Astrophysics},
  volume={57},
  number={1},
  pages={617--663},
  year={2019},
  publisher={Annual Reviews}
}

@article{delamer2024,
  title={TOI-4201: An Early M Dwarf Hosting a Massive Transiting Jupiter Stretching Theories of Core Accretion},
  author={Delamer, Megan and Kanodia, Shubham and Ca{\~n}as, Caleb I and M{\"u}ller, Simon and Helled, Ravit and Lin, Andrea SJ and Libby-Roberts, Jessica E and Gupta, Arvind F and Mahadevan, Suvrath and Teske, Johanna and others},
  journal={The Astrophysical Journal Letters},
  volume={962},
  number={2},
  pages={L22},
  year={2024},
  publisher={IOP Publishing}
}

@article{andrews2013,
  title={The mass dependence between protoplanetary disks and their stellar hosts},
  author={Andrews, Sean M and Rosenfeld, Katherine A and Kraus, Adam L and Wilner, David J},
  journal={The Astrophysical Journal},
  volume={771},
  number={2},
  pages={129},
  year={2013},
  publisher={IOP Publishing}
}

@article{pascucci2016,
  title={A steeper than linear disk mass--stellar mass scaling relation},
  author={Pascucci, Ilaria and Testi, Leonardo and Herczeg, Gregory J and Long, F and Manara, CF and Hendler, N and Mulders, Gijs D and Krijt, S and Ciesla, F and Henning, Th and others},
  journal={The Astrophysical Journal},
  volume={831},
  number={2},
  pages={125},
  year={2016},
  publisher={IOP Publishing}
}

@article{stevenson1982,
  title={Interiors of the giant planets},
  author={Stevenson, David J},
  journal={In: Annual review of earth and planetary sciences. Volume 10.(A82-35776 17-88) Palo Alto, CA, Annual Reviews, Inc., 1982, p. 257-295.},
  volume={10},
  pages={257--295},
  year={1982}
}

@article{pollack1996,
  title={Formation of the giant planets by concurrent accretion of solids and gas},
  author={Pollack, James B and Hubickyj, Olenka and Bodenheimer, Peter and Lissauer, Jack J and Podolak, Morris and Greenzweig, Yuval},
  journal={icarus},
  volume={124},
  number={1},
  pages={62--85},
  year={1996},
  publisher={Elsevier}
}

@article{chachan2023,
  title={Small planets around cool dwarfs: enhanced formation efficiency of super-Earths around M dwarfs},
  author={Chachan, Yayaati and Lee, Eve J},
  journal={The Astrophysical Journal Letters},
  volume={952},
  number={1},
  pages={L20},
  year={2023},
  publisher={IOP Publishing}
}

@article{boss2023,
  title={Forming gas giants around a range of protostellar M-dwarfs by gas disk gravitational instability},
  author={Boss, Alan P and Kanodia, Shubham},
  journal={The Astrophysical Journal},
  volume={956},
  number={1},
  pages={4},
  year={2023},
  publisher={IOP Publishing}
}

@article{tychoniec2020,
  title={Dust masses of young disks: constraining the initial solid reservoir for planet formation},
  author={Tychoniec, {\L}ukasz and Manara, Carlo F and Rosotti, Giovanni P and van Dishoeck, Ewine F and Cridland, Alexander J and Hsieh, Tien-Hao and Murillo, Nadia M and Segura-Cox, Dominique and van Terwisga, Sierk E and Tobin, John J},
  journal={Astronomy \& Astrophysics},
  volume={640},
  pages={A19},
  year={2020},
  publisher={EDP Sciences}
}

@article{ribas2023,
  title={The CARMENES search for exoplanets around M dwarfs-Guaranteed time observations Data Release 1 (2016-2020)},
  author={Ribas, I and Reiners, A and Zechmeister, M and Caballero, JA and Morales, JC and Sabotta, S and Baroch, D and Amado, PJ and Quirrenbach, A and Abril, M and others},
  journal={Astronomy \& Astrophysics},
  volume={670},
  pages={A139},
  year={2023},
  publisher={EDP Sciences}
}

@article{bonfils2013,
  title={The HARPS search for southern extra-solar planets-XXXI. The M-dwarf sample},
  author={Bonfils, Xavier and Delfosse, Xl and Udry, S and Forveille, T and Mayor, M and Perrier, C and Bouchy, Fran{\c{c}}ois and Gillon, Micha{\"e}l and Lovis, C and Pepe, F and others},
  journal={Astronomy \& Astrophysics},
  volume={549},
  pages={A109},
  year={2013},
  publisher={EDP Sciences}
}

@article{pinamonti2022,
  title={HADES RV Programme with HARPS-N at TNG-XV. Planetary occurrence rates around early-M dwarfs},
  author={Pinamonti, Matteo and Sozzetti, Alessandro and Maldonado, Jes{\'u}s and Affer, L and Micela, Giusi and Bonomo, AS and Lanza, AF and Perger, M and Ribas, Ignasi and Hern{\'a}ndez, JI Gonz{\'a}lez and others},
  journal={Astronomy \& Astrophysics},
  volume={664},
  pages={A65},
  year={2022},
  publisher={EDP Sciences}
}

@article{pass2023,
  title={Mid-to-late M dwarfs lack Jupiter analogs},
  author={Pass, Emily K and Winters, Jennifer G and Charbonneau, David and Irwin, Jonathan M and Latham, David W and Berlind, Perry and Calkins, Michael L and Esquerdo, Gilbert A and Mink, Jessica},
  journal={The Astronomical Journal},
  volume={166},
  number={1},
  pages={11},
  year={2023},
  publisher={IOP Publishing}
}

@article{mignon2025,
  title={Radial velocity homogeneous analysis of M dwarfs observed with HARPS. II. Detection limits and planetary occurrence statistics},
  author={Mignon, L and Delfosse, X and Meunier, N and Chaverot, G and Burn, R and Bonfils, X and Bouchy, F and Astudillo-Defru, N and Curto, G Lo and Gaisne, G and others},
  journal={arXiv preprint arXiv:2502.06553},
  year={2025}
}

@article{alibert2005,
  title={Models of giant planet formation with migration and disc evolution},
  author={Alibert, Yann and Mordasini, Christoph and Benz, Willy and Winisdoerffer, Christophe},
  journal={Astronomy \& Astrophysics},
  volume={434},
  number={1},
  pages={343--353},
  year={2005},
  publisher={EDP Sciences}
}

@article{liu2020,
  title={A tale of planet formation: from dust to planets},
  author={Liu, Beibei and Ji, Jianghui},
  journal={Research in Astronomy and Astrophysics},
  volume={20},
  number={10},
  pages={164},
  year={2020},
  publisher={IOP Publishing}
}

@article{mercer2020,
  title={Planet formation around M dwarfs via disc instability-Fragmentation conditions and protoplanet properties},
  author={Mercer, Anthony and Stamatellos, Dimitris},
  journal={Astronomy \& Astrophysics},
  volume={633},
  pages={A116},
  year={2020},
  publisher={EDP Sciences}
}

@article{somers2020spots,
  title={The spots models: a grid of theoretical stellar evolution tracks and isochrones for testing the effects of starspots on structure and colors},
  author={Somers, Garrett and Cao, Lyra and Pinsonneault, Marc H},
  journal={The Astrophysical Journal},
  volume={891},
  number={1},
  pages={29},
  year={2020},
  publisher={IOP Publishing}
}

@article{allard2013bt,
  title={The BT-settl model atmospheres for stars, brown dwarfs and planets},
  author={Allard, F},
  journal={Proceedings of the International Astronomical Union},
  volume={8},
  number={S299},
  pages={271--272},
  year={2013},
  publisher={Cambridge University Press}
}

@article{rackham2024toward,
  title={Toward Robust Corrections for Stellar Contamination in JWST Exoplanet Transmission Spectra},
  author={Rackham, Benjamin V and de Wit, Julien},
  journal={The Astronomical Journal},
  volume={168},
  number={2},
  pages={82},
  year={2024},
  publisher={IOP Publishing}
}

@article{lim2023atmospheric,
  title={Atmospheric reconnaissance of TRAPPIST-1 b with JWST/NIRISS: evidence for strong stellar contamination in the transmission spectra},
  author={Lim, Olivia and Benneke, Bj{\"o}rn and Doyon, Ren{\'e} and MacDonald, Ryan J and Piaulet, Caroline and Artigau, {\'E}tienne and Coulombe, Louis-Philippe and Radica, Michael and L’Heureux, Alexandrine and Albert, Lo{\"\i}c and others},
  journal={The Astrophysical Journal Letters},
  volume={955},
  number={1},
  pages={L22},
  year={2023},
  publisher={IOP Publishing}
}

@article{garcia2022hst,
  title={HST/WFC3 transmission spectroscopy of the cold rocky planet TRAPPIST-1h},
  author={Garcia, LJ and Moran, SE and Rackham, BV and Wakeford, HR and Gillon, Micha{\"e}l and de Wit, Julien and Lewis, NK},
  journal={Astronomy \& Astrophysics},
  volume={665},
  pages={A19},
  year={2022},
  publisher={EDP Sciences}
}

@article{rafikov2005can,
  title={Can giant planets form by direct gravitational instability?},
  author={Rafikov, Roman R},
  journal={The Astrophysical Journal},
  volume={621},
  number={1},
  pages={L69},
  year={2005},
  publisher={IOP Publishing}
}

@article{andrews2007high,
  title={High-resolution submillimeter constraints on circumstellar disk structure},
  author={Andrews, Sean M and Williams, Jonathan P},
  journal={The Astrophysical Journal},
  volume={659},
  number={1},
  pages={705},
  year={2007},
  publisher={IOP Publishing}
}

@article{masset2006,
  title={Disk surface density transitions as protoplanet traps},
  author={Masset, FS and Morbidelli, A and Crida, A and Ferreira, Jorge},
  journal={The Astrophysical Journal},
  volume={642},
  number={1},
  pages={478},
  year={2006},
  publisher={IOP Publishing}
}

@article{kratter2010,
  title={The runts of the litter: why planets formed through gravitational instability can only be failed binary stars},
  author={Kratter, Kaitlin M and Murray-Clay, Ruth A and Youdin, Andrew N},
  journal={The Astrophysical Journal},
  volume={710},
  number={2},
  pages={1375},
  year={2010},
  publisher={IOP Publishing}
}

@article{speagle2020dynesty,
  title={dynesty: a dynamic nested sampling package for estimating Bayesian posteriors and evidences},
  author={Speagle, Joshua S},
  journal={Monthly Notices of the Royal Astronomical Society},
  volume={493},
  number={3},
  pages={3132--3158},
  year={2020},
  publisher={Oxford University Press}
}

@article{rotman2025enabling,
  title={Enabling Robust Exoplanet Atmospheric Retrievals with Gaussian Processes},
  author={Rotman, Yoav and Welbanks, Luis and Line, Michael R and McGill, Peter and Radica, Michael and Nixon, Matthew C},
  journal={arXiv preprint arXiv:2503.21702},
  year={2025}
}

@article{yurchenko2020exomol,
  title={ExoMol line lists--XXXIX. Ro-vibrational molecular line list for CO2},
  author={Yurchenko, SN and Mellor, Thomas M and Freedman, Richard S and Tennyson, J},
  journal={Monthly Notices of the Royal Astronomical Society},
  volume={496},
  number={4},
  pages={5282--5291},
  year={2020},
  publisher={Oxford University Press}
}

@article{yurchenko2024exomol,
  title={ExoMol line lists--LVII. High accuracy ro-vibrational line list for methane (CH4)},
  author={Yurchenko, Sergei N and Owens, Alec and Kefala, Kyriaki and Tennyson, Jonathan},
  journal={Monthly Notices of the Royal Astronomical Society},
  volume={528},
  number={2},
  pages={3719--3729},
  year={2024},
  publisher={Oxford University Press}
}

@article{li2015rovibrational,
  title={Rovibrational line lists for nine isotopologues of the CO molecule in the X1$\Sigma$+ ground electronic state},
  author={Li, Gang and Gordon, Iouli E and Rothman, Laurence S and Tan, Yan and Hu, Shui-Ming and Kassi, Samir and Campargue, Alain and Medvedev, Emile S},
  journal={The Astrophysical Journal Supplement Series},
  volume={216},
  number={1},
  pages={15},
  year={2015},
  publisher={IOP Publishing}
}

@article{coles2019exomol,
  title={ExoMol molecular line lists--XXXV. A rotation-vibration line list for hot ammonia},
  author={Coles, Phillip A and Yurchenko, Sergei N and Tennyson, Jonathan},
  journal={Monthly Notices of the Royal Astronomical Society},
  volume={490},
  number={4},
  pages={4638--4647},
  year={2019},
  publisher={Oxford University Press}
}

@article{barber2014exomol,
  title={ExoMol line lists--III. An improved hot rotation-vibration line list for HCN and HNC},
  author={Barber, RJ and Strange, JK and Hill, C and Polyansky, OL and Mellau, G Ch and Yurchenko, SN and Tennyson, Jonathan},
  journal={Monthly Notices of the Royal Astronomical Society},
  volume={437},
  number={2},
  pages={1828--1835},
  year={2014},
  publisher={Oxford University Press}
}

@article{polyansky2018exomol,
  title={ExoMol molecular line lists XXX: a complete high-accuracy line list for water},
  author={Polyansky, Oleg L and Kyuberis, Aleksandra A and Zobov, Nikolai F and Tennyson, Jonathan and Yurchenko, Sergei N and Lodi, Lorenzo},
  journal={Monthly Notices of the Royal Astronomical Society},
  volume={480},
  number={2},
  pages={2597--2608},
  year={2018},
  publisher={Oxford University Press}
}

@article{kawashima2019,
  title={Theoretical Transmission Spectra of Exoplanet Atmospheres with Hydrocarbon Haze: Effect of Creation, Growth, and Settling of Haze Particles. II. Dependence on UV Irradiation Intensity, Metallicity, C/O Ratio, Eddy Diffusion Coefficient, and Temperature (vol 877, 109, 2019)},
  author={Kawashima, Yui and Ikoma, Masahiro},
  journal={ASTROPHYSICAL JOURNAL},
  volume={884},
  number={1},
  year={2019},
  publisher={IOP PUBLISHING LTD TEMPLE CIRCUS, TEMPLE WAY, BRISTOL BS1 6BE, ENGLAND}
}

@misc{gao2021,
  title={Aerosols in exoplanet atmospheres},
  author={Gao, Peter and Wakeford, Hannah R and Moran, Sarah E and Parmentier, Vivien},
  year={2021},
  publisher={Wiley Online Library}
}

@article{sing2016,
  title={A continuum from clear to cloudy hot-Jupiter exoplanets without primordial water depletion},
  author={Sing, David K and Fortney, Jonathan J and Nikolov, Nikolay and Wakeford, Hannah R and Kataria, Tiffany and Evans, Thomas M and Aigrain, Suzanne and Ballester, Gilda E and Burrows, Adam S and Deming, Drake and others},
  journal={Nature},
  volume={529},
  number={7584},
  pages={59--62},
  year={2016},
  publisher={Nature Publishing Group UK London}
}

@article{sedaghati2017,
  title={Detection of titanium oxide in the atmosphere of a hot Jupiter},
  author={Sedaghati, Elyar and Boffin, Henri MJ and MacDonald, Ryan J and Gandhi, Siddharth and Madhusudhan, Nikku and Gibson, Neale P and Oshagh, Mahmoudreza and Claret, Antonio and Rauer, Heike},
  journal={Nature},
  volume={549},
  number={7671},
  pages={238--241},
  year={2017},
  publisher={Nature Publishing Group UK London}
}

@article{may2019,
  title={MOPSS. II. Extreme optical scattering slope for the inflated super-Neptune HATS-8b},
  author={May, EM and Gardner, T and Rauscher, E and Monnier, JD},
  journal={The Astronomical Journal},
  volume={159},
  number={1},
  pages={7},
  year={2019},
  publisher={IOP Publishing}
}

@article{ohno2020,
  title={Super-Rayleigh slopes in transmission spectra of exoplanets generated by photochemical haze},
  author={Ohno, Kazumasa and Kawashima, Yui},
  journal={The Astrophysical Journal Letters},
  volume={895},
  number={2},
  pages={L47},
  year={2020},
  publisher={IOP Publishing}
}

@article{espinoza2019,
  title={ACCESS: a featureless optical transmission spectrum for WASP-19b from Magellan/IMACS},
  author={Espinoza, N{\'e}stor and Rackham, Benjamin V and Jord{\'a}n, Andr{\'e}s and Apai, D{\'a}niel and L{\'o}pez-Morales, Mercedes and Osip, David J and Grimm, Simon L and Hoeijmakers, Jens and Wilson, Paul A and Bixel, Alex and others},
  journal={Monthly Notices of the Royal Astronomical Society},
  volume={482},
  number={2},
  pages={2065--2087},
  year={2019},
  publisher={Oxford University Press}
}

@article{molliere2022,
  title={Interpreting the atmospheric composition of exoplanets: sensitivity to planet formation assumptions},
  author={Molli{\`e}re, Paul and Molyarova, Tamara and Bitsch, Bertram and Henning, Thomas and Schneider, Aaron and Kreidberg, Laura and Eistrup, Christian and Burn, Remo and Nasedkin, Evert and Semenov, Dmitry and others},
  journal={The Astrophysical Journal},
  volume={934},
  number={1},
  pages={74},
  year={2022},
  publisher={IOP Publishing}
}

@article{mulders2021,
  title={The mass budgets and spatial scales of exoplanet systems and protoplanetary disks},
  author={Mulders, Gijs D and Pascucci, Ilaria and Ciesla, Fred J and Fernandes, Rachel B},
  journal={The Astrophysical Journal},
  volume={920},
  number={2},
  pages={66},
  year={2021},
  publisher={IOP Publishing}
}

@article{manara2022,
  title={Demographics of young stars and their protoplanetary disks: lessons learned on disk evolution and its connection to planet formation},
  author={Manara, Carlo F and Ansdell, Megan and Rosotti, Giovanni P and Hughes, A Meredith and Armitage, Philip J and Lodato, Giuseppe and Williams, Jonathan P},
  journal={arXiv preprint arXiv:2203.09930},
  year={2022}
}

@article{schib2025,
  title={DIPSY: A new Disc Instability Population SYnthesis-II. The Populations of Companions Formed Through Disc Instability},
  author={Schib, O and Mordasini, C and Emsenhuber, A and Helled, R},
  journal={Astronomy \& Astrophysics},
  volume={704},
  pages={A28},
  year={2025},
  publisher={EDP Sciences}
}

@article{mordasini2024,
  title={Planet formation—Observational constraints, physical processes, and compositional patterns},
  author={Mordasini, Christoph and Burn, Remo},
  journal={Reviews in Mineralogy and Geochemistry},
  volume={90},
  number={1},
  pages={55--112},
  year={2024},
  publisher={Mineralogical Society of America}
}

@article{vulcan,
  title={VULCAN: an open-source, validated chemical kinetics Python code for exoplanetary atmospheres},
  author={Tsai, Shang-Min and Lyons, James R and Grosheintz, Luc and Rimmer, Paul B and Kitzmann, Daniel and Heng, Kevin},
  journal={The Astrophysical Journal Supplement Series},
  volume={228},
  number={2},
  pages={20},
  year={2017},
  publisher={IOP Publishing}
}

@article{underwood2016exomol,
  title={ExoMol molecular line lists--XIV. The rotation--vibration spectrum of hot SO2},
  author={Underwood, Daniel S and Tennyson, Jonathan and Yurchenko, Sergei N and Huang, Xinchuan and Schwenke, David W and Lee, Timothy J and Clausen, S{\o}nnik and Fateev, Alexander},
  journal={Monthly Notices of the Royal Astronomical Society},
  volume={459},
  number={4},
  pages={3890--3899},
  year={2016},
  publisher={Oxford University Press}
}

@article{azzam2016exomol,
  title={ExoMol molecular line lists--XVI. The rotation--vibration spectrum of hot H2S},
  author={Azzam, Ala'a AA and Tennyson, Jonathan and Yurchenko, Sergei N and Naumenko, Olga V},
  journal={Monthly Notices of the Royal Astronomical Society},
  volume={460},
  number={4},
  pages={4063--4074},
  year={2016},
  publisher={Oxford University Press}
}

@article{chubb2020exomol,
  title={ExoMol molecular line lists--XXXVII. Spectra of acetylene},
  author={Chubb, Katy L and Tennyson, Jonathan and Yurchenko, Sergei N},
  journal={Monthly Notices of the Royal Astronomical Society},
  volume={493},
  number={2},
  pages={1531--1545},
  year={2020},
  publisher={Oxford University Press}
}

@article{gordon2022hitran,
  title={The HITRAN2020 molecular spectroscopic database},
  author={Gordon, Iouli E and Rothman, Laurence S and Hargreaves, ea RJ and Hashemi, R and Karlovets, Ekaterina Vladimirovna and Skinner, FM and Conway, Eamon K and Hill, Christian and Kochanov, Roman V and Tan, Y and others},
  journal={Journal of quantitative spectroscopy and radiative transfer},
  volume={277},
  pages={107949},
  year={2022},
  publisher={Elsevier}
}

@article{fu2024,
  title={Hydrogen sulfide and metal-enriched atmosphere for a Jupiter-mass exoplanet},
  author={Fu, Guangwei and Welbanks, Luis and Deming, Drake and Inglis, Julie and Zhang, Michael and Lothringer, Joshua and Ih, Jegug and Moses, Julianne I and Schlawin, Everett and Knutson, Heather A and others},
  journal={Nature},
  volume={632},
  number={8026},
  pages={752--756},
  year={2024},
  publisher={Nature Publishing Group UK London}
}

@article{tsai2023,
  title={Photochemically produced SO2 in the atmosphere of WASP-39b},
  author={Tsai, Shang-Min and Lee, Elspeth KH and Powell, Diana and Gao, Peter and Zhang, Xi and Moses, Julianne and H{\'e}brard, Eric and Venot, Olivia and Parmentier, Vivien and Jordan, Sean and others},
  journal={Nature},
  volume={617},
  number={7961},
  pages={483--487},
  year={2023},
  publisher={Nature Publishing Group UK London}
}

@article{crossfield2025,
  title={Mapping the SO2 Shoreline in Gas Giant Exoplanets},
  author={Crossfield, Ian JM and Ahrer, Eva-Maria and Brande, Jonathan and Kreidberg, Laura and Lothringer, Joshua and Piaulet-Ghorayeb, Caroline and Polman, Jesse and Welbanks, Luis and Kirk, James and Powell, Diana and others},
  journal={The Astrophysical Journal},
  volume={994},
  number={2},
  pages={184},
  year={2025},
  publisher={The American Astronomical Society}
}

@article{madhusudhan2012,
  title={C/O ratio as a dimension for characterizing exoplanetary atmospheres},
  author={Madhusudhan, Nikku},
  journal={The Astrophysical Journal},
  volume={758},
  number={1},
  pages={36},
  year={2012},
  publisher={The American Astronomical Society}
}

@article{zahnle2009,
  title={Atmospheric sulfur photochemistry on hot Jupiters},
  author={Zahnle, Kevin and Marley, Mark S and Freedman, R St and Lodders, K and Fortney, JJ},
  journal={The Astrophysical Journal},
  volume={701},
  number={1},
  pages={L20--L24},
  year={2009},
  publisher={The American Astronomical Society}
}

@article{heng2016,
  title={Analytical Models of Exoplanetary Atmospheres. III. Gaseous CHON Chemistry with 9 Molecules},
  author={Heng, Kevin and Tsai, Shang-Min},
  journal={arXiv preprint arXiv:1603.05418},
  year={2016}
}

@article{lim2015pysynphot,
  title={Pysynphot user’s guide},
  author={Lim, PL and Diaz, RI and Laidler, V},
  journal={STScI, Baltimore, MD},
  year={2015}
}

@article{macdonald2017,
  title={HD 209458b in new light: evidence of nitrogen chemistry, patchy clouds and sub-solar water},
  author={MacDonald, Ryan J and Madhusudhan, Nikku},
  journal={Monthly Notices of the Royal Astronomical Society},
  volume={469},
  number={2},
  pages={1979--1996},
  year={2017},
  publisher={Oxford University Press}
}
\bibliographystyle{aasjournal}

\appendix
\section{Chemical Equilibrium Predictions} \label{app:A}
To place our retrieved molecular abundances in a physically motivated context, we compute chemical equilibrium abundance predictions for TOI-3235~b using \texttt{VULCAN} \citep{vulcan}. 

\begin{figure*}
    \centering
    \includegraphics[width=0.9\linewidth]{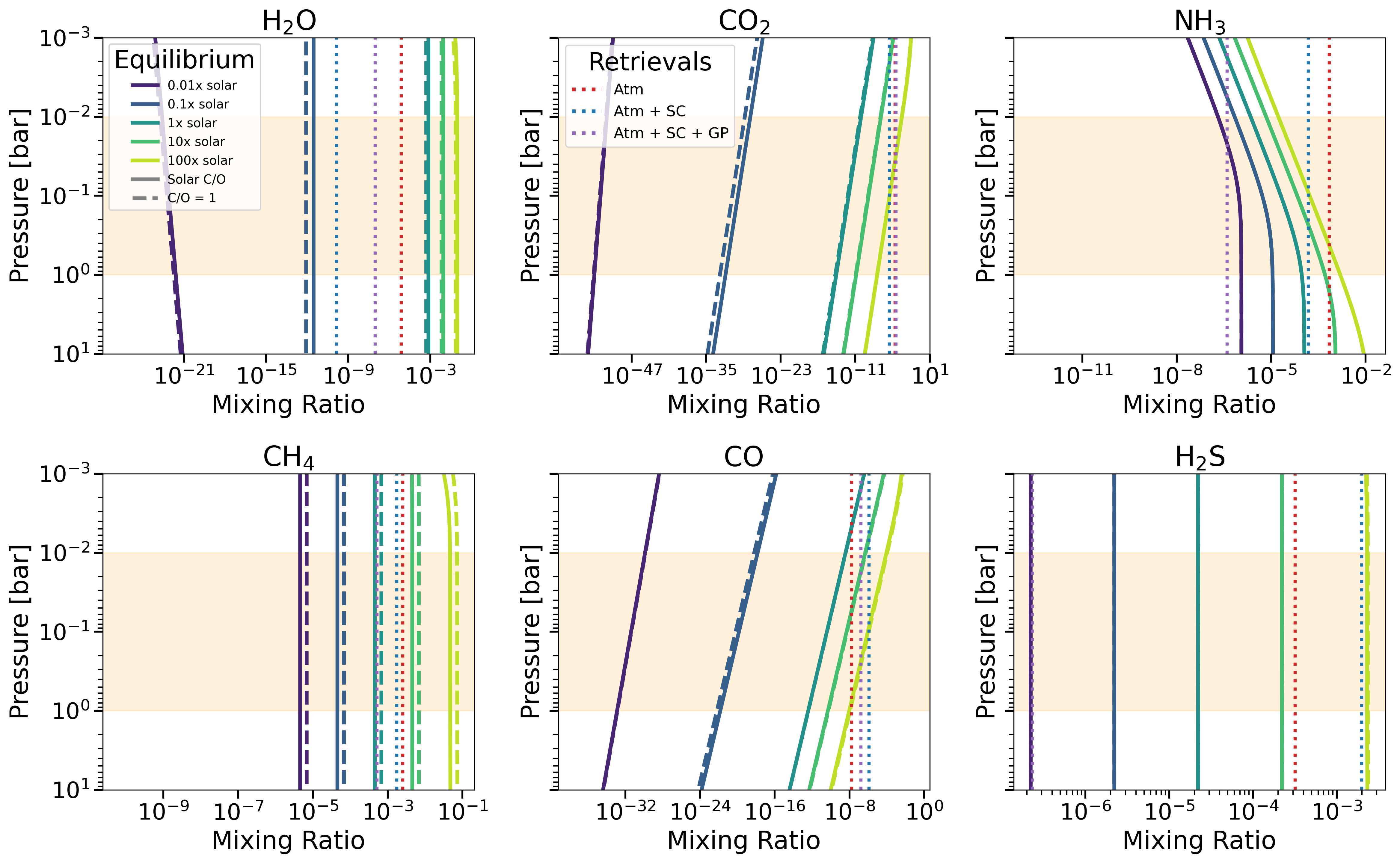}
    \caption{
    Chemical equilibrium abundance profiles computed with \texttt{VULCAN} for TOI-3235~b across metallicities from $0.01\times$ to $100\times$ solar. Solid colored curves show models with solar C/O (0.55), while dashed colored curves show models with super-solar C/O (1). The shaded orange region marks the approximate pressure range probed by the transmission spectrum, $10^{-2}$--$10$~bar, estimated from pressure contribution functions. Solid colored curves show the equilibrium predictions, while dashed vertical lines show the median retrieved abundances from the atmosphere-only, atm+SC, and atm+SC+GP retrievals. The six species shown are those whose equilibrium abundances overlap our retrieval prior range of $10^{-12}$--$10^{-1}$ within this pressure region.
    }
    \label{fig:vulcan_eq}
\end{figure*}

We adopt $T_{\rm eq}=604$~K and $\log g = 3.2$, and explore metallicities from $0.01\times$ to $100\times$ solar for both solar C/O (0.55) and super-solar C/O (1). Figure~\ref{fig:vulcan_eq} compares these predictions to the median abundances from our three retrievals for the species whose equilibrium abundances overlap our retrieval prior range within the approximate transmission-probed pressure range.

Within this pressure range, H$_2$O, CO$_2$, and CO fall below the lower bound of our retrieval prior at low metallicities, so we summarize them by their maximum predicted abundances across the grid: $\log X_{\rm H_2O} \lesssim -1.03$, $\log X_{\rm CO_2} \lesssim -3.47$, and $\log X_{\rm CO} \lesssim -4.03$. The remaining species have equilibrium abundance ranges of $\log X_{\rm NH_3} = -6.69$ to $-2.06$, $\log X_{\rm CH_4} = -5.35$ to $-1.32$, and $\log X_{\rm H_2S} = -6.66$ to $-2.63$.

\section{Posterior Distributions} \label{app:B}

\begin{figure*}
    \centering
    \includegraphics[width=0.9\linewidth]{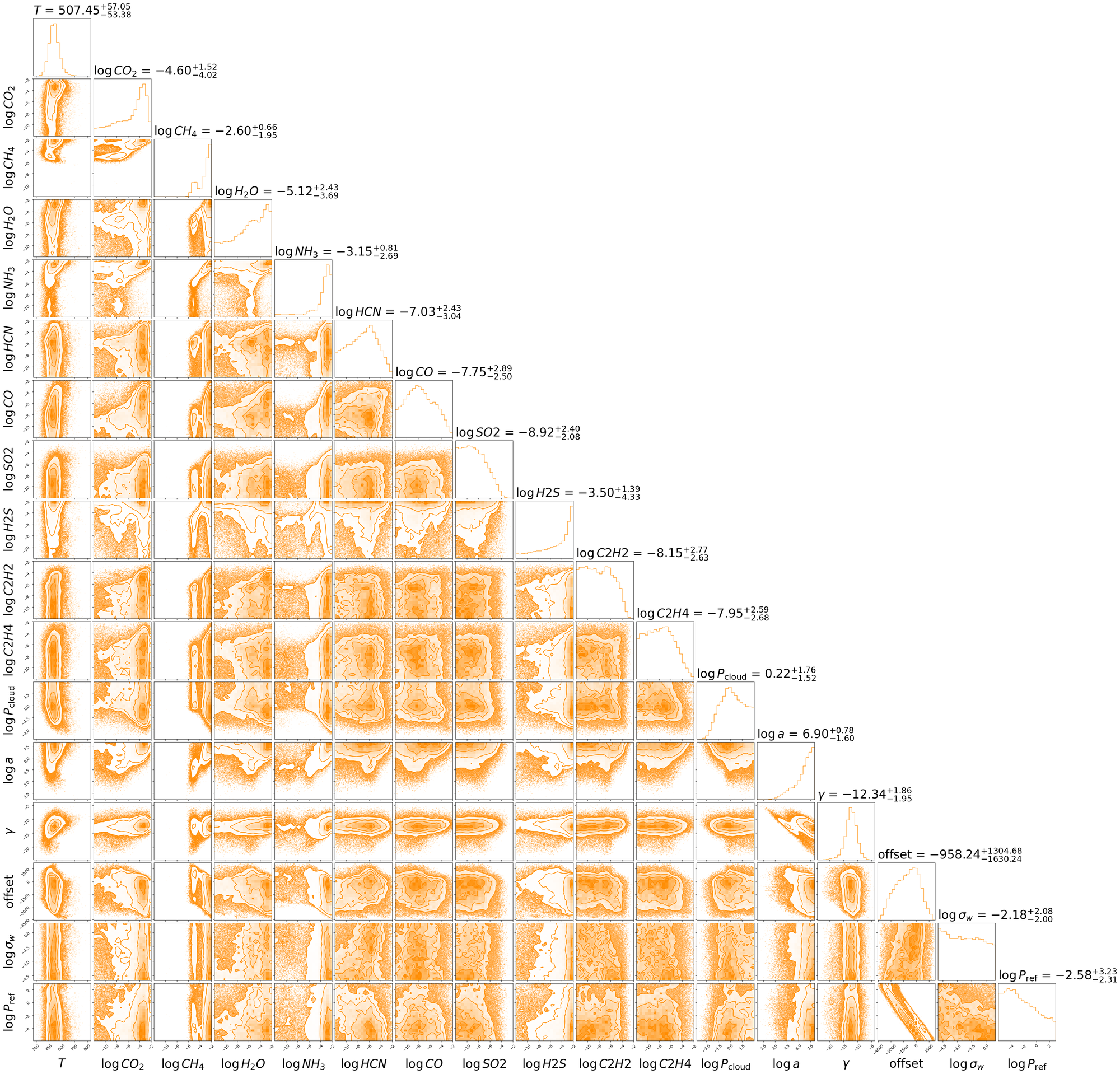}
    \caption{Posterior distributions from retrieval fitting for the atmospheric model.}
    \label{fig:atm_cp}
\end{figure*}

\begin{figure*}
    \centering
    \includegraphics[width=0.9\linewidth]{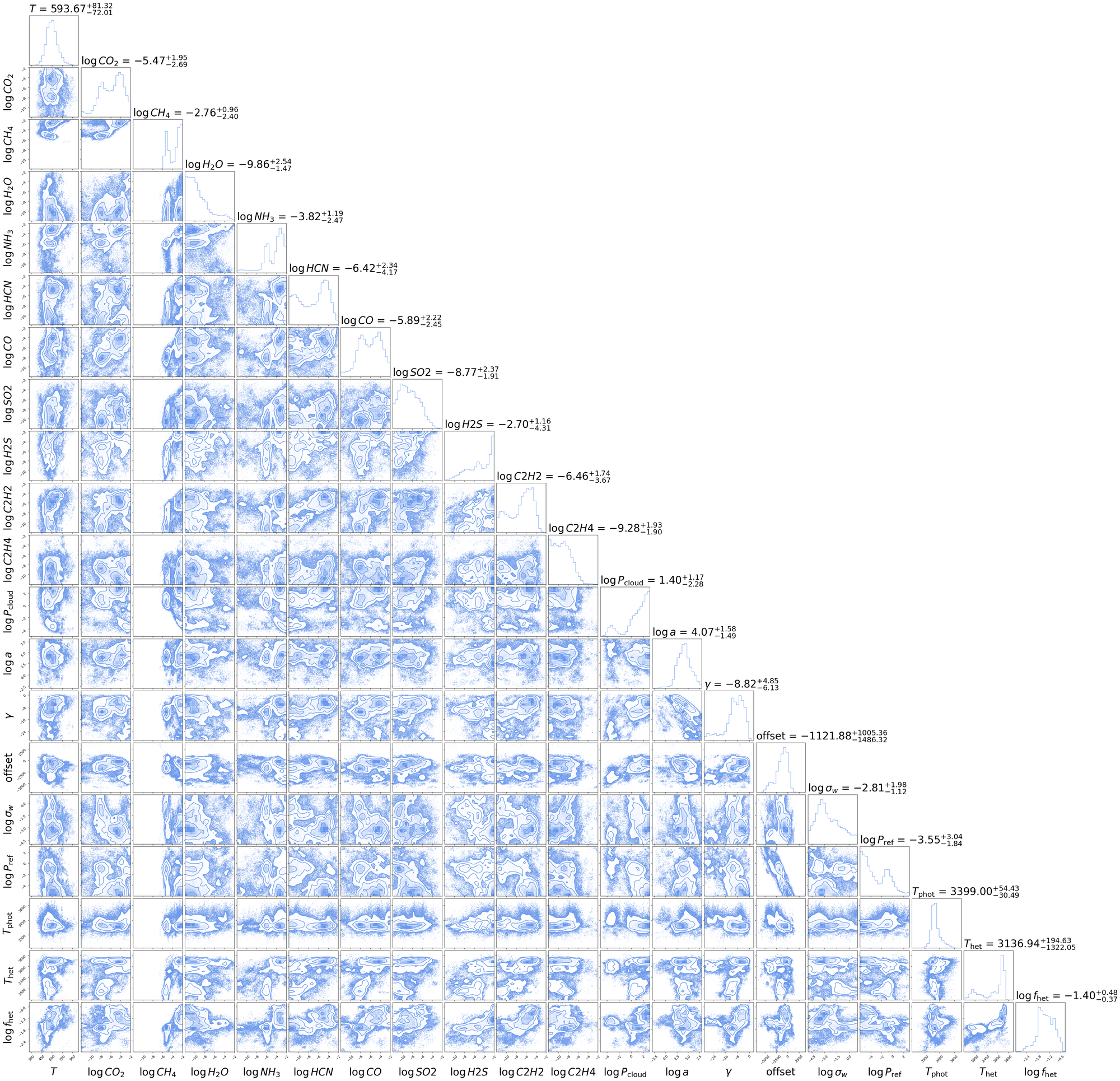}
    \caption{Posterior distributions from retrieval fitting for the atmospheric model and stellar contamination model.}
    \label{fig:atm_sc_cp}
\end{figure*}

\begin{figure*}
    \centering
    \includegraphics[width=0.9\linewidth]{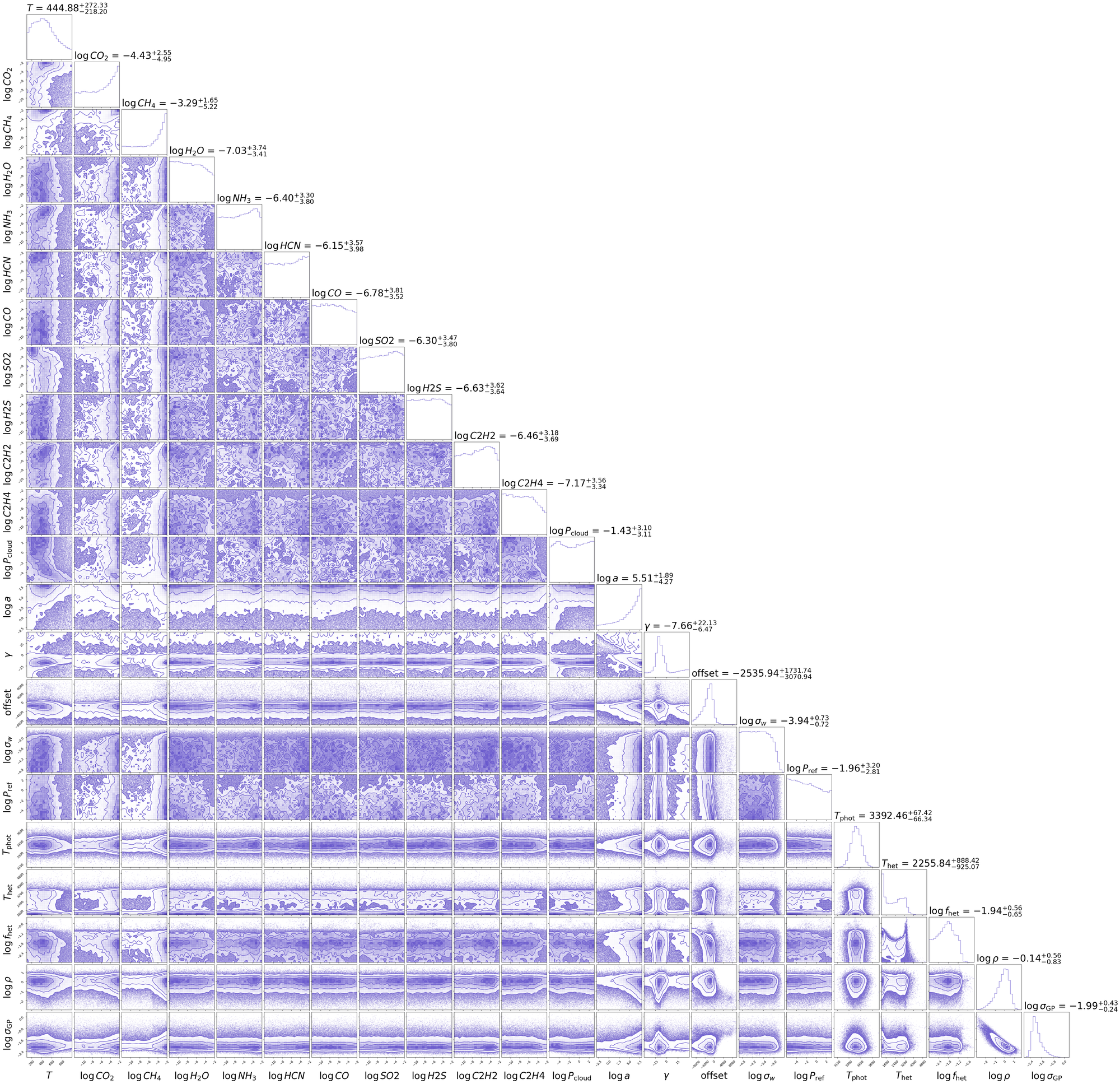}
    \caption{Posterior distributions from retrieval fitting for the atmospheric model, the stellar contamination model, and a Gaussian process.}
    \label{fig:atm_sc_gp_cp}
\end{figure*}

Figures~\ref{fig:atm_cp} to \ref{fig:atm_sc_gp_cp} show the posterior distributions for the three retrieval configurations discussed in Section~\ref{sec:results}: the atmosphere-only model, the atmosphere plus stellar contamination model, and the atmosphere plus stellar contamination plus GP model.

\section{Retrievals on Truncated and Untruncated Spectrum} \label{app:C}
\begin{figure*}
    \centering
    \includegraphics[width=0.9\linewidth]{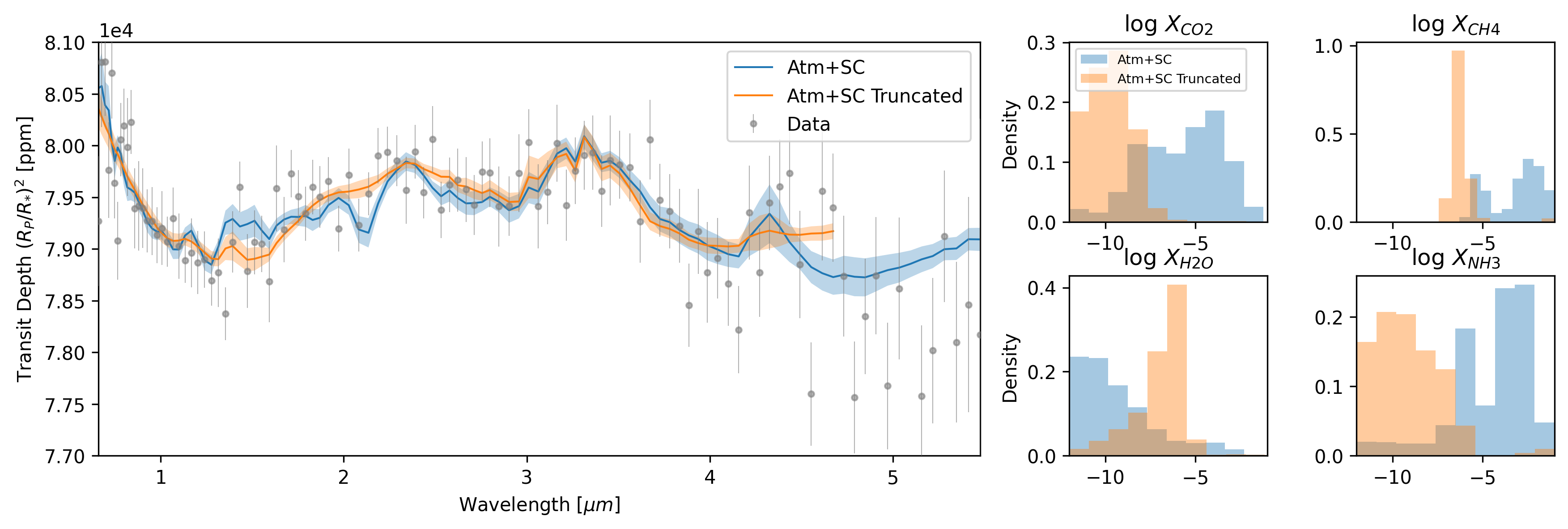}
    \caption{Atm+SC retrieval on the full spectrum and the truncated spectrum. The left panel shows both median retrieved spectra and 1$\sigma$ bounds and the right shows the retrieved abundances for four trace species. The two retrievals yield different results.}
    \label{fig:atm_sc_trunc}
\end{figure*}

\begin{figure*}
    \centering
    \includegraphics[width=0.9\linewidth]{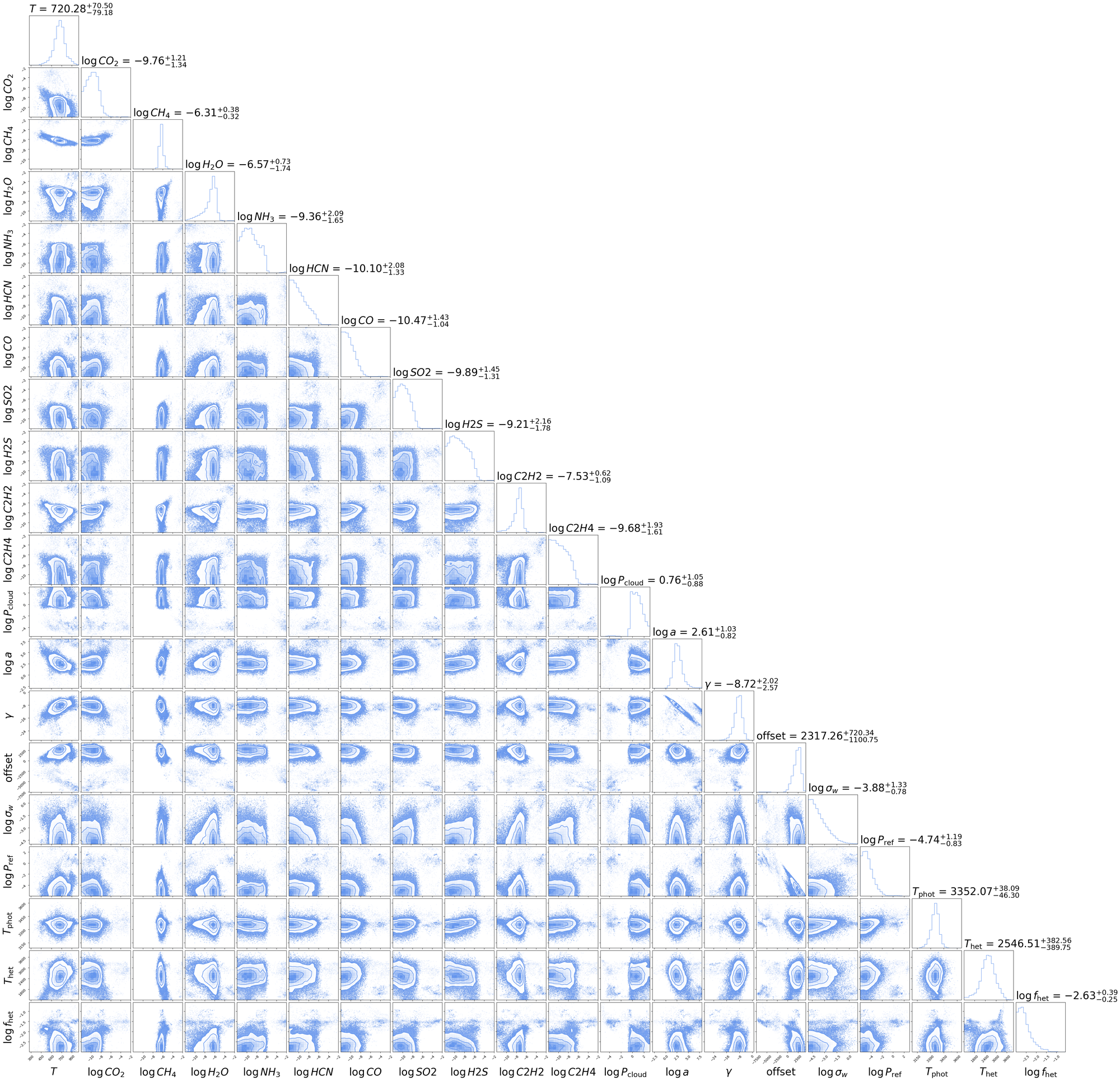}
    \caption{Posterior distributions from atm+SC retrieval on the truncated spectrum.}
    \label{fig:atm_sc_trunc_posteriors}
\end{figure*}

\begin{figure*}
    \centering
    \includegraphics[width=0.9\linewidth]{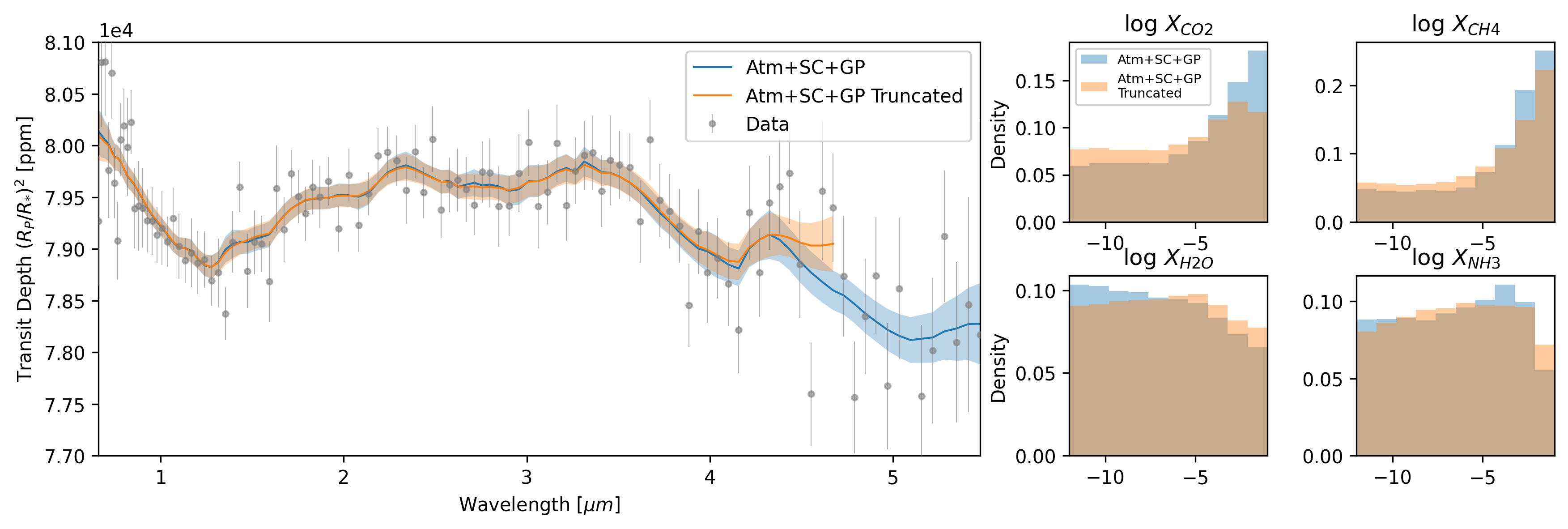}
    \caption{Atm+SC+GP retrieval on the full spectrum and the truncated spectrum. The left panel shows both median retrieved spectra and 1$\sigma$ bounds and the right shows the retrieved abundances for four trace species. The two retrievals yield similar results.}
    \label{fig:atm_sc_gp_trunc}
\end{figure*}

\begin{figure*}
    \centering
    \includegraphics[width=0.9\linewidth]{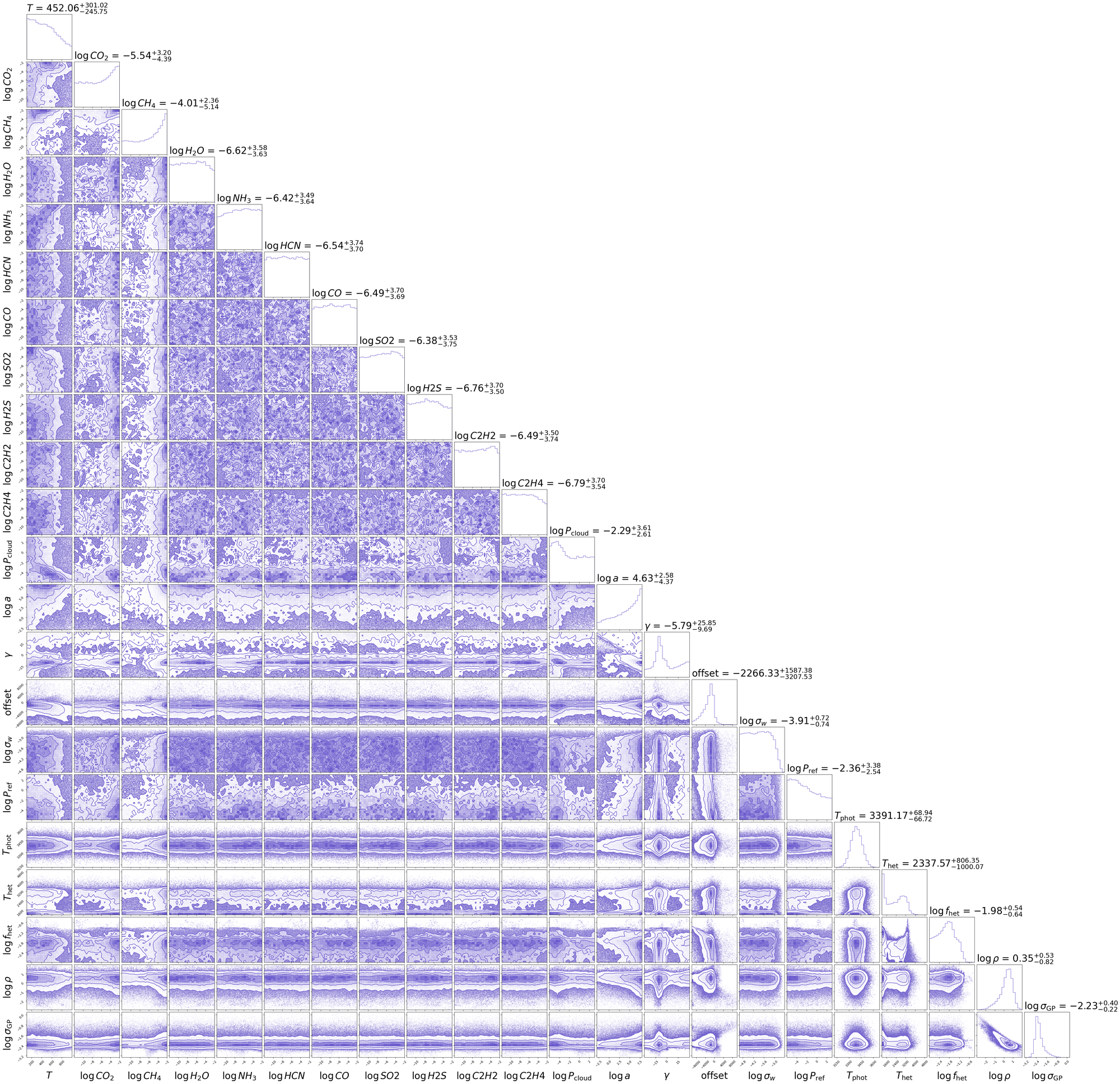}
    \caption{Posterior distributions from atm+SC+GP retrieval on the truncated spectrum.}
    \label{fig:atm_sc_gp_trunc_posteriors}
\end{figure*}

Figures~\ref{fig:atm_sc_trunc} to \ref{fig:atm_sc_gp_trunc_posteriors} compare retrievals performed on the full transmission spectrum and on a spectrum truncated at $4.7~\mu$m, showing how the long-wavelength structure affects the inferred atmospheric properties with and without the GP component.

\section{Atmospheric Metallicity and C/O Prior Correction} \label{app:D}

In our free-chemistry retrievals, individual molecular abundances are assigned independent log-uniform priors. Atmospheric metallicity and C/O are then computed as derived quantities from the abundance samples. These derived quantities therefore inherit non-uniform induced priors from the abundance parameterization.

\begin{figure}
    \centering
    \includegraphics[width=0.8\textwidth]{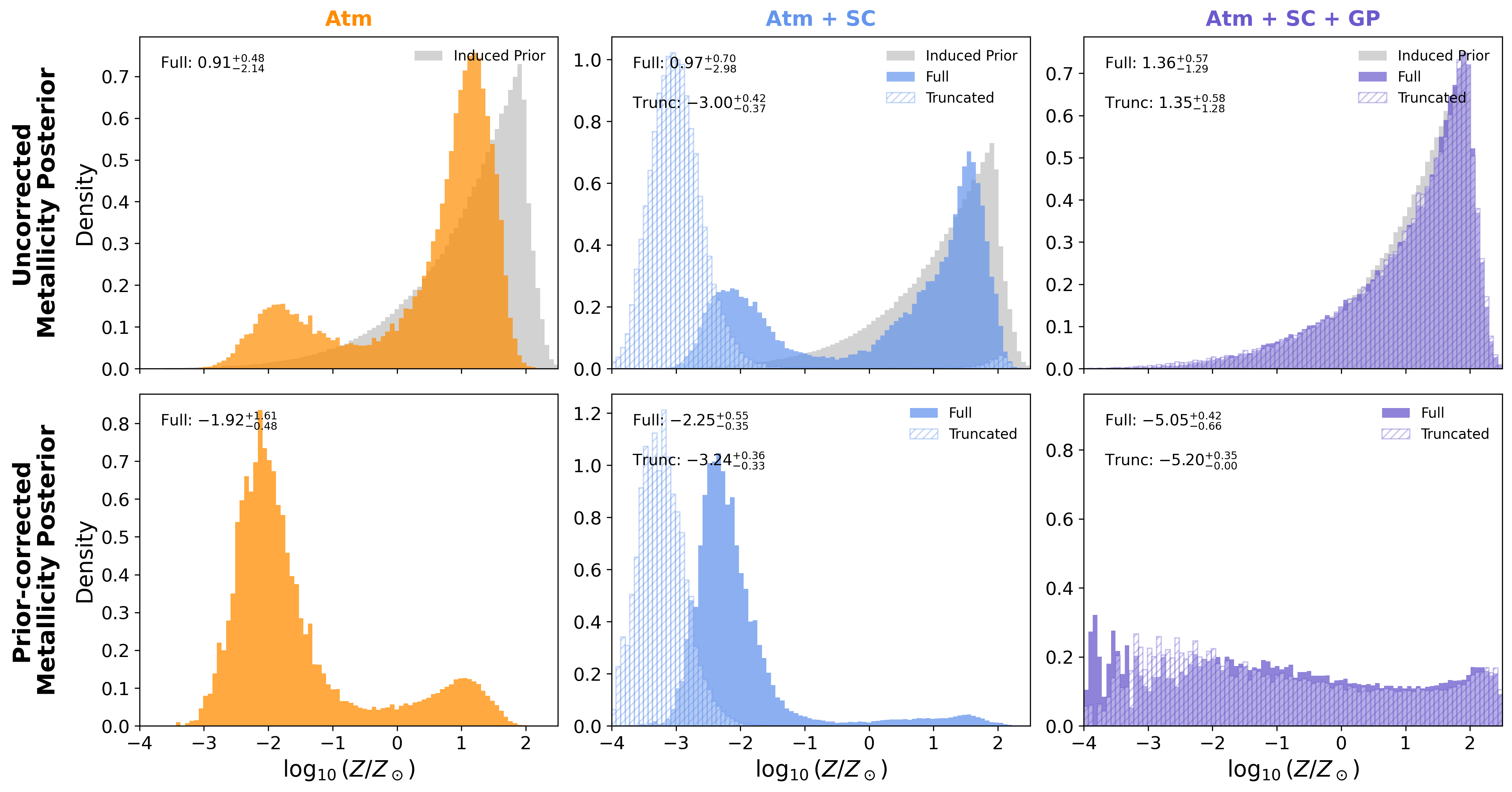}
    \caption{
    Derived metallicity posteriors before and after correcting for the induced metallicity prior. The top row shows the uncorrected metallicity posteriors, with the induced metallicity prior shown in gray. The bottom row shows the prior-corrected posteriors obtained using the inverse-density weights from Equation~\ref{eq:prior_reweight}. The black vertical line marks the solar metallicity value, and colored vertical lines mark the posterior medians.
    }
    \label{fig:corrected_metallicities}
\end{figure}

\begin{figure}
    \centering
    \includegraphics[width=0.8\textwidth]{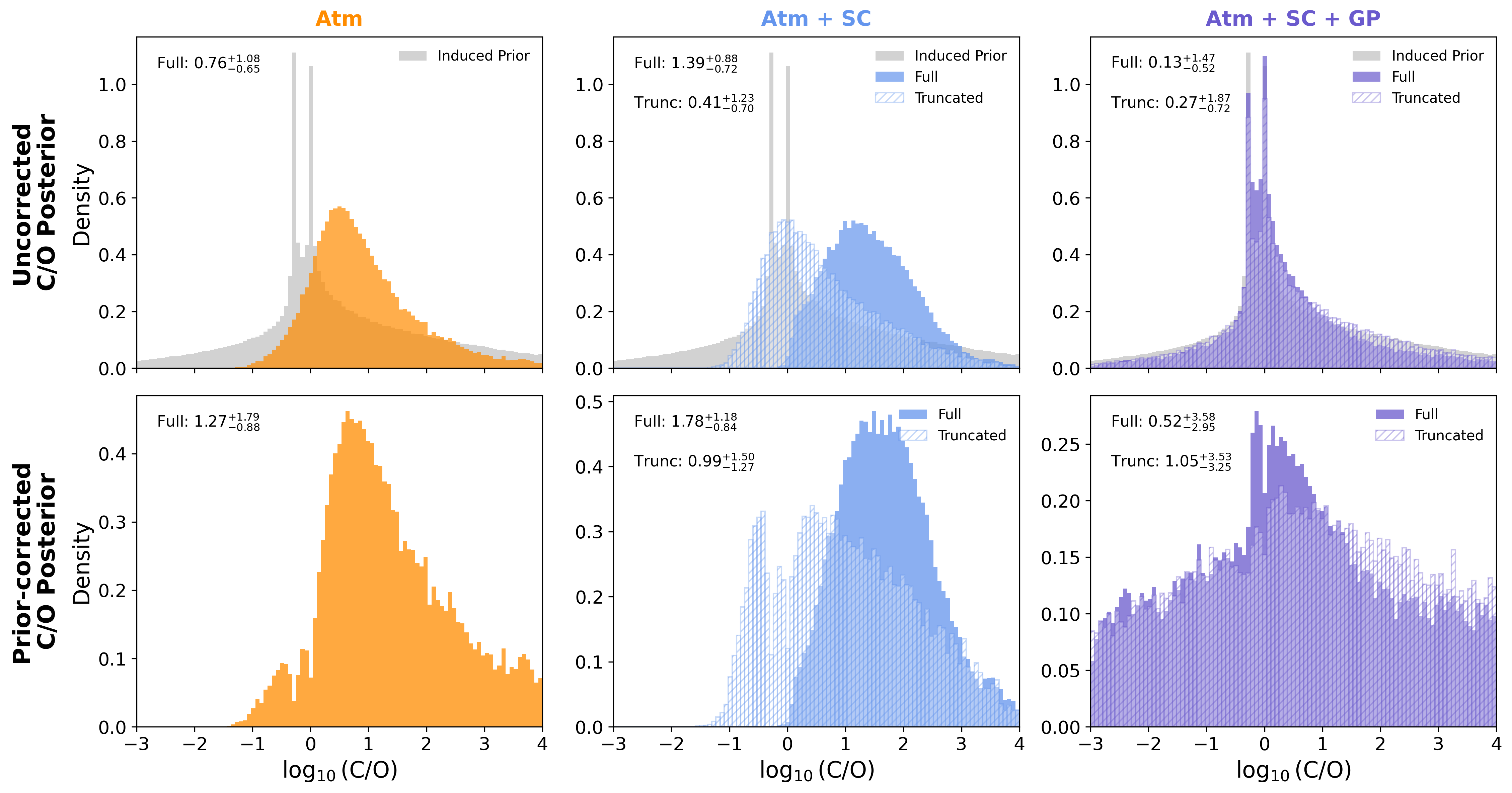}
    \caption{
    Derived C/O posteriors before and after correcting for the induced C/O prior. The top row shows uncorrected posteriors in $\log_{10}({\rm C/O})$, with the induced C/O prior shown in gray. The bottom row shows prior-corrected posteriors obtained using the inverse-density weights from Equation~\ref{eq:prior_reweight}, corresponding to a target prior uniform in $\log_{10}({\rm C/O})$. The black vertical line marks the solar C/O ratio, and colored vertical lines mark the posterior medians.
    }
    \label{fig:corrected_co}
\end{figure}

We estimate the induced priors by drawing $5\times10^5$ samples from the same molecular abundance priors used in the retrievals and computing $\log_{10}(Z/Z_\odot)$ and $\log_{10}({\rm C/O})$ for each draw. These induced priors are shown as gray histograms in Figures~\ref{fig:corrected_metallicities} and \ref{fig:corrected_co}.

We correct these induced prior-volume effects using importance reweighting. For each derived quantity, we estimate the prior density implied by our abundance priors using the prior samples described above. Each posterior sample is then assigned a weight
\begin{equation}
    w_j \propto \frac{1}{\pi_{\rm ind}(q_j)},
    \label{eq:prior_reweight}
\end{equation}
where $q_j$ is the value of the derived quantity for the $j$th posterior sample and $\pi_{\rm ind}$ is the induced prior density evaluated at that value. This is equivalent to adopting a target prior that is uniform in the chosen derived quantity.

For metallicity, we take $q=\log_{10}(Z/Z_\odot)$, so the corrected posterior corresponds to a target prior uniform in $\log_{10}(Z/Z_\odot)$. For C/O, we take $q=\log_{10}({\rm C/O})$, corresponding to a target prior uniform in $\log_{10}({\rm C/O})$. We choose this log-uniform C/O target prior because the free-abundance priors allow C/O to span several orders of magnitude, especially when oxygen-bearing abundances are small.

Figure~\ref{fig:corrected_metallicities} compares the original and prior-corrected metallicity posteriors. The induced metallicity prior is strongly weighted toward super-solar values. The correction shifts the atm and atm+SC metallicity estimates to sub-solar values, whereas the atm+SC+GP case broadens, indicating that the metallicity is weakly constrained once the GP component is included. The prior-corrected atm+SC+GP posterior exhibits some noise at low metallicity because this region is sparsely sampled under the induced prior, and the few posterior samples there receive large inverse-prior weights. This sampling noise does not affect our conclusion: the atm+SC+GP retrieval does not provide strong evidence for a narrowly constrained super-solar metallicity but rather a broad metallicity posterior.

Figure~\ref{fig:corrected_co} shows the analogous comparison for C/O. The C/O prior is structured, with features near the stoichiometric ratios of individual molecules, such as C/O $\simeq 0.5$ and C/O $\simeq 1$, reflecting the C/O ratios of CO$_2$ and CO, respectively. The atm and atm+SC posteriors remain shifted toward high C/O after reweighting, while the atm+SC+GP posterior broadens. The prior-corrected atm+SC+GP C/O posterior exhibits some noise in the low- and high-C/O wings because these regions are sparsely sampled under the induced prior, and the few posterior samples there receive large inverse-prior weights. This sampling noise does not affect our conclusion: the atm+SC+GP retrieval does not provide strong evidence for a narrowly constrained C/O ratio, but rather yields a broad C/O posterior once the GP component is included.

These prior corrections are important for correctly interpreting the derived atmospheric properties. For metallicity, the correction shifts the atmosphere-only and atm+SC posteriors toward lower metallicities, while for C/O it does not qualitatively change the interpretation that these two retrievals favor super-solar values. However, both corrections are crucial for the atm+SC+GP retrieval. Without reweighting, the atm+SC+GP metallicity and C/O posteriors would appear artificially well constrained. The prior-corrected distributions reveal that this apparent constraint is largely inherited from the induced priors rather than being data-driven. After correction, both posteriors broaden substantially, showing that the GP properly propagates uncertainty from residual wavelength-correlated structure into the inferred metallicity and C/O. This demonstrates that prior reweighting is necessary to avoid over-interpreting prior-dominated derived quantities.

\section{GP Synthetic Retrieval Test} \label{app:gp_injection}

To test whether the inclusion of a GP absorbs all physical atmospheric and stellar contamination features, we perform a synthetic retrieval test. We generate a synthetic transmission spectrum from a representative atm+SC model using the median parameters from the atm+SC retrieval as described in Section~\ref{subsec:SC_results}. We then add Gaussian noise using the same uncertainties as the observed TOI-3235~b transmission spectrum. This synthetic dataset therefore represents a case where the deterministic atm+SC model is sufficient to model the data.

\begin{figure*}
    \centering
    \includegraphics[width=\linewidth]{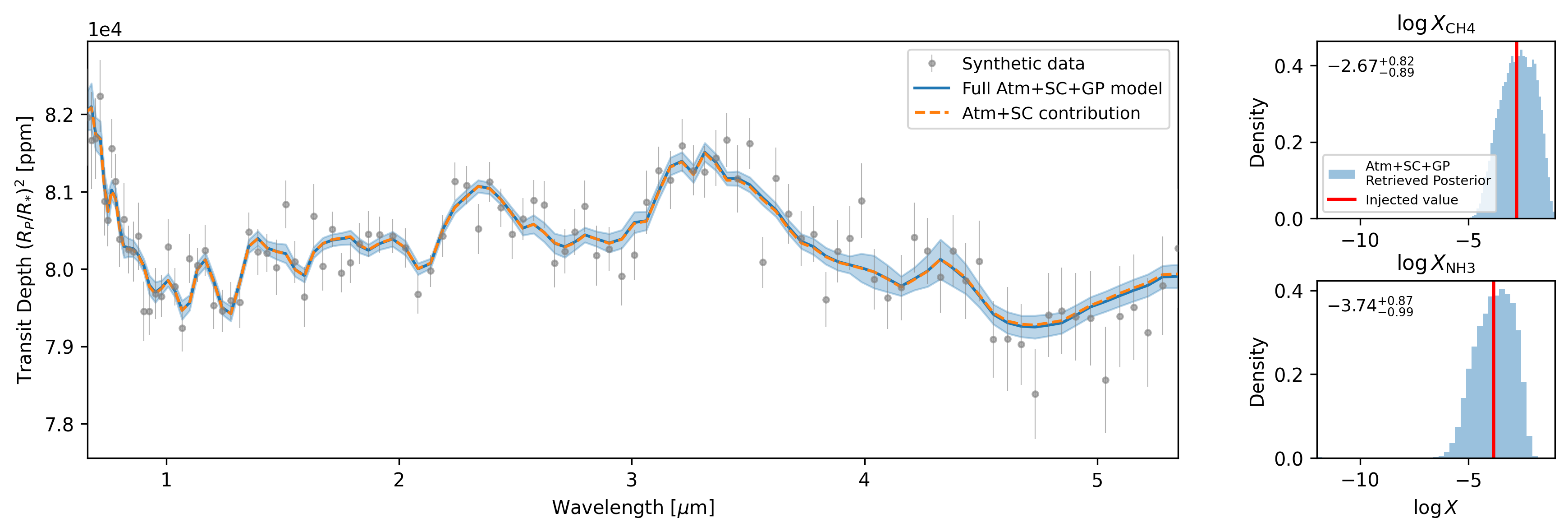}
    \caption{
    Synthetic retrieval test assessing whether the GP component absorbs atmospheric and stellar contamination signals. The synthetic data were generated from a representative atm+SC model using the median parameters from the atm+SC retrieval, with Gaussian noise added following the observed transmission spectrum uncertainties. The blue curve and shaded region show the median and 1$\sigma$ interval of the full atm+SC+GP retrieved spectrum. The orange dashed curve shows the deterministic atm+SC contribution from the same retrieval with the GP component removed for visualization. The consistency between the blue and orange curves shows that the GP does not dominate the fit when the deterministic model is sufficient to explain the data. The right panels show the retrieved CH$_4$ and NH$_3$ abundance posteriors, with red vertical lines marking the injected values.
    }
    \label{fig:gp_synthetic_retrieval}
\end{figure*}

We retrieve this synthetic spectrum using the atm+SC+GP framework as described in Section~\ref{subsec:GP_setup}. Figure~\ref{fig:gp_synthetic_retrieval} shows the result. The blue curve and shaded region show the full atm+SC+GP retrieved spectrum and its 1$\sigma$ interval, while the orange dashed curve shows the deterministic atm+SC contribution from the same retrieval with the GP component removed for visualization. The consistency between these two curves indicates that, when the deterministic model can describe the data, the GP does not dominate the fit. Instead, the spectral structure is recovered by the deterministic atm+SC model component.

The right panels of Figure~\ref{fig:gp_synthetic_retrieval} show the retrieved abundance posteriors for CH$_4$ and NH$_3$, two of the best-constrained species in this synthetic retrieval. The red vertical lines mark the injected values, which are recovered within the 1$\sigma$ posterior intervals. This test demonstrates that, given our retrieval framework and data quality, including a GP does not automatically erase atmospheric or stellar contamination information when the deterministic components are adequate descriptions of the data. The broad posteriors obtained for the real TOI-3235~b spectrum with the inclusion of a GP can therefore be interpreted as arising from wavelength-correlated structure not fully captured by the deterministic atm+SC model, rather than as an unavoidable consequence of including a GP.

\end{document}